\documentclass{iopart}

\pdfoutput=1

\usepackage{iopams}
\usepackage{amssymb}
\usepackage{graphicx}
\usepackage{mathrsfs}
\usepackage{epstopdf}

\def\RnR{rock'n'roller}
\def\frak#1#2{\textstyle\frac{#1}{#2}}
\def\half{\textstyle\frac{1}{2}}
\def\quarter{\textstyle\frac{1}{4}}

\def\IA{I_1} 
\def\IB{I_2} 
\def\IC{I_3}

\def\bomdot{\mathbf{\dot{\bomega}}}
\def\bthdot{\mathbf{\dot{\btheta}}}

\renewcommand{\bdelta}{\boldsymbol{\delta}}

\begin{document}

\title{Quaternion Solution for the Rock'n'roller: \\
       Box Orbits, Loop Orbits and Recession}

\author{Peter Lynch and Miguel D Bustamante}

\address{School of Mathematical Sciences,
        UCD, Belfield, Dublin 4, Ireland}
\ead{
     \mailto{Peter.Lynch@ucd.ie},
     \mailto{Miguel.Bustamante@ucd.ie}}


\submitto{R\&CD \hfill \bf Draft: 28 June 2012}



\begin{abstract}

We consider two types of trajectories found in a wide range of
mechanical systems, \emph{viz.} box orbits and loop orbits. We
elucidate the dynamics of these orbits in the simple context of
a perturbed harmonic oscillator in two dimensions.
We then examine the small-amplitude motion of a rigid body, the
rock'n'roller, a sphere with eccentric distribution of mass.
The equations of motion are expressed in quaternionic form and
a complete analytical solution is obtained. Both types of orbit,
boxes and loops, are found, the particular form depending on the
initial conditions. We interpret the motion in terms of epi-elliptic
orbits.  The phenomenon of \emph{recession}, or reversal of
precession, is associated with box orbits. The small-amplitude
solutions for the symmetric case, or Routh sphere, are expressed
explicitly in terms of epicycles; there is no recession in this case.

\end{abstract}


\section{Introduction: Box Orbits and Loop Orbits}
\label{sec:Introduction}

\subsection{Libration and rotation}
\label{sec:LibRot}

The simple pendulum, with one degree of freedom, provides a
valuable model for a wide range of physical phenomena.
The pendulum is constrained to move in a plane, and has two
essentially different modes of behaviour.
In \emph{libration}, the bob oscillates about the suspension
point, and the angular momentum reverses sign periodically.
In \emph{rotation} the bob moves in a circle with the angular momentum
varying periodically but remaining always of one sign. In
many systems with more than one degree of freedom there are analogues of
these two distinct behaviour patterns. In \S\ref{sec:PSHO}
we investigate this distinction for a perturbed simple harmonic
oscillator in two dimensions.  In \S\ref{sec:RnR-Eqns}
we investigate the dynamics of the \RnR\ and derive a complete
solution for small amplitude motions in terms of quaternions.
The dynamics in the case of small asymmetry, 
$\epsilon=(\IB-\IA)/\IA\ll 1$, are examined in \S\ref{sec:RnRasym}.
The special case of a symmetric body, the Routh Sphere ($\epsilon=0$),
is considered in \S\ref{sec:RouthSphere} and the
solutions are expressed explicitly in terms of epicycles.
Concluding remarks are made in \S\ref{sec:Conclusion}.


\subsection{Stelar motion in a globular cluster}
\label{sec:GlobularClusters}

\begin{figure}
\begin{center}
\includegraphics[scale=0.25]{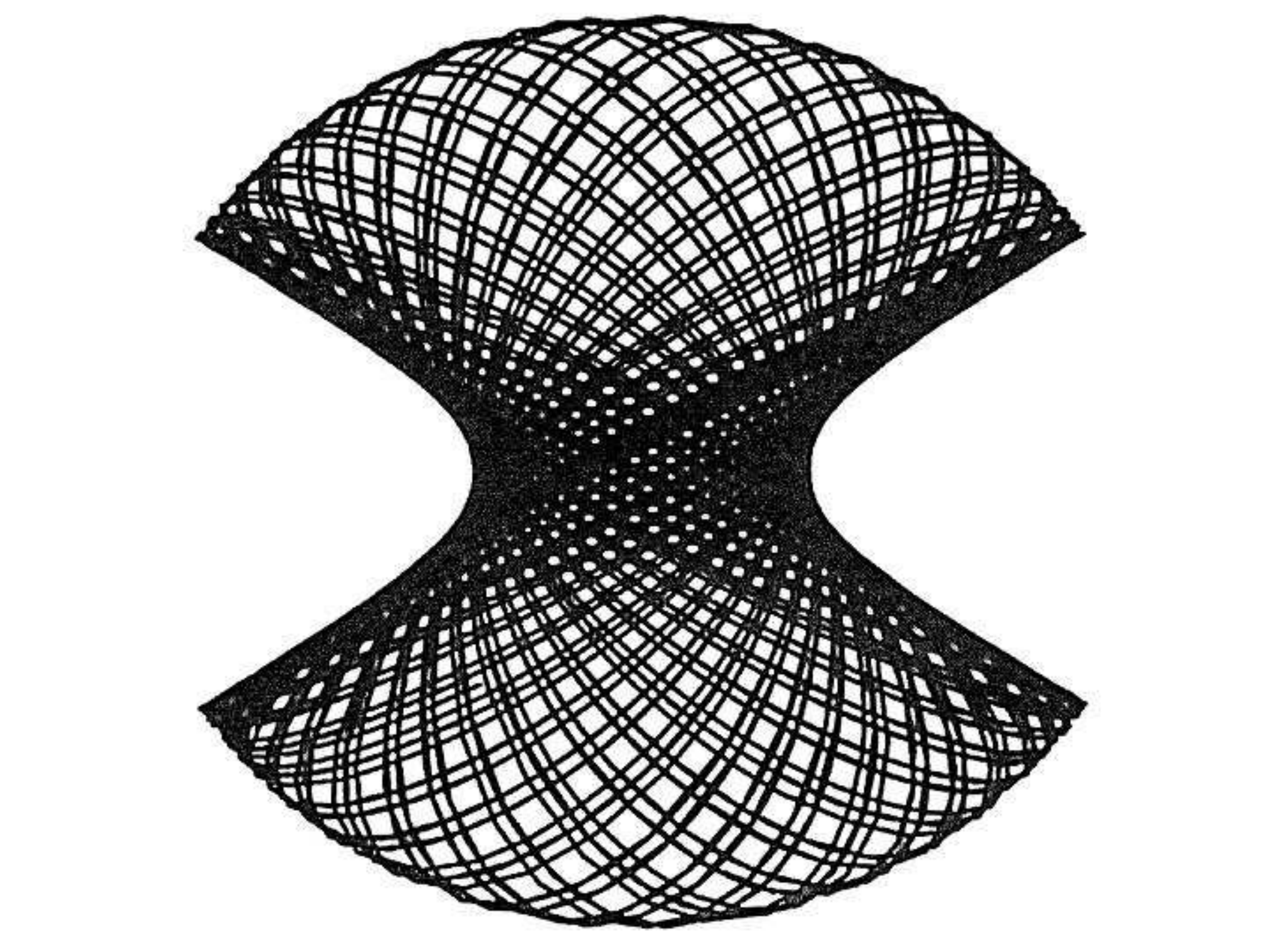}
\includegraphics[scale=0.25]{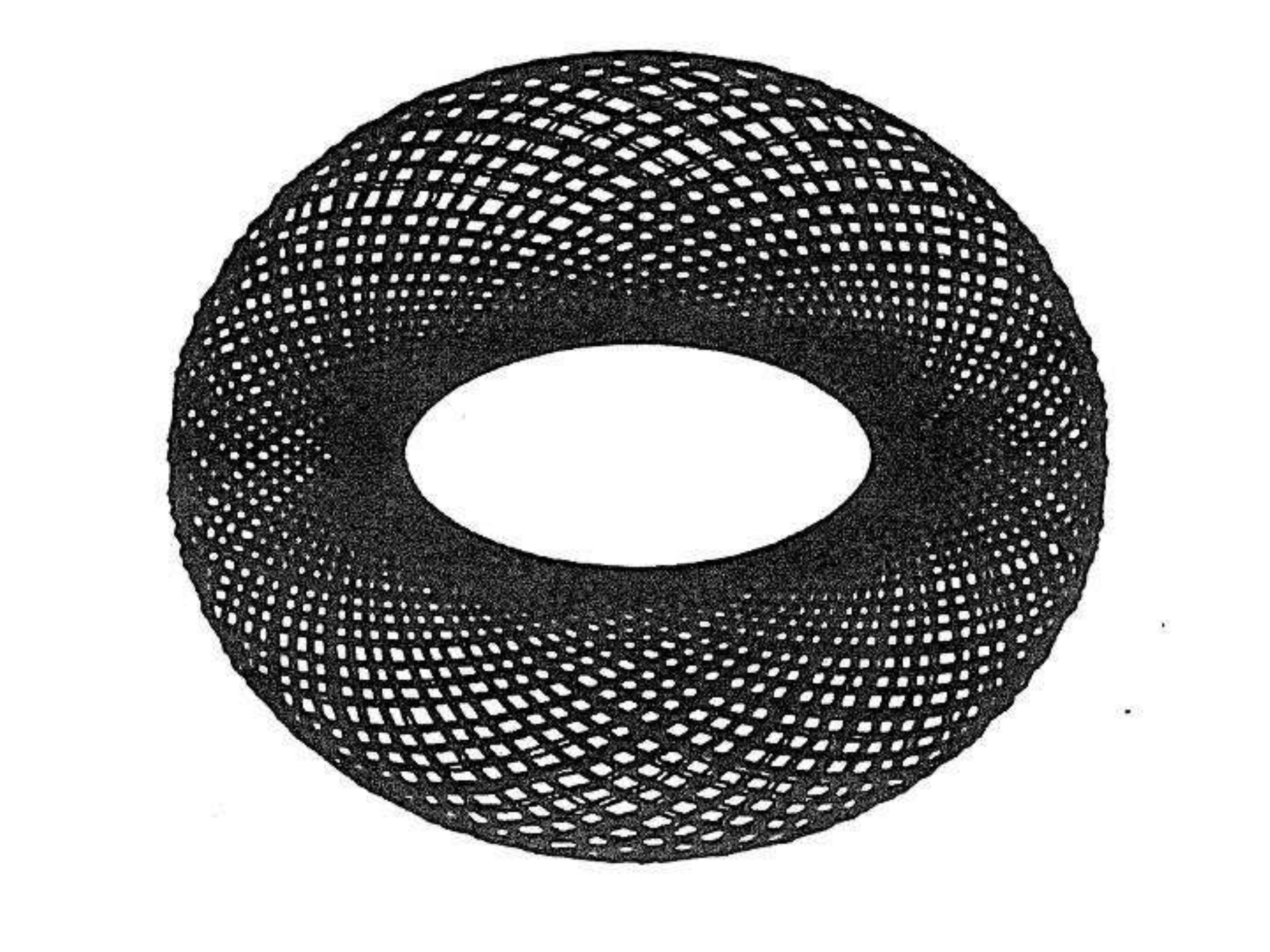}
\caption{Two orbits in a logarithmic gravitational potential.
Left: a box orbit.  Right: a loop orbit.
Both have equal energy and the character of the
orbit is determined by the initial conditions. Figure taken from
\cite[pg.~174]{BT2008}, where further details may be found.}
\label{fig:GlobClus}
\end{center}
\end{figure}

In stellar systems such as triaxial globular clusters,
which do not have symmetry about any of the three axes, two distinct
types of orbit are found.
Since the force is not central, the angular momentum is not conserved.
If we consider motions in the symmetry plane perpendicular to one axis,
with differing frequencies about the other two directions,
we can distinguish two possibilities.
In a \emph{box orbit}, a star oscillates independently
about the two axes as it moves along its orbit.
As a result of this motion, it fills in a simply connected region
of space that includes the centre and that, for small amplitude,
approximates a rectangle.
The star is free to come arbitrarily close to the centre of the
system. If the frequencies with respect to the axes are
rationally related, the orbit will be closed. It will then resemble a 
Lissajous curve. 
The angular momentum takes both positive and negative values.
In a \emph{loop orbit}, the angular momentum about a perpendicular
to the orbital plane remains of one sign.  The orbit fills a region
limited by two approximately elliptic curves,
and is bounded away from the centre.
We illustrate the two orbit types in Fig~\ref{fig:GlobClus},
taken from \cite[pg.~174]{BT2008}.

\subsection{Motions of the \RnR}
\label{sec:Rocknroller}

\begin{figure}
\begin{center}
\includegraphics[scale=0.67]{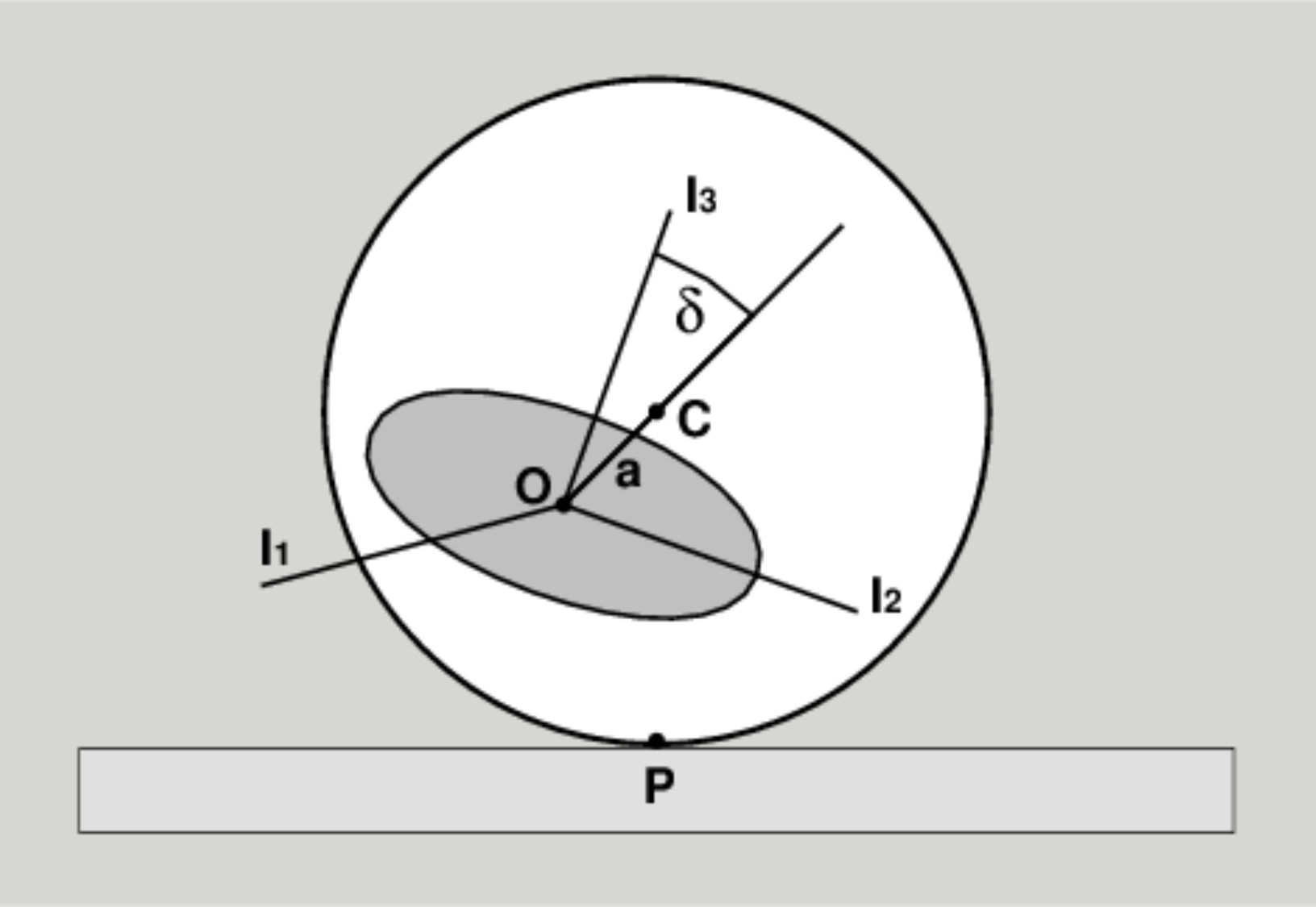}

\caption
{The loaded sphere. In this model, all the mass is contained
in the triaxial ellipsoid (shaded), with moments of inertia
$\IA\le\IB\le\IC$ along the principal axes at the centre of mass $O$.
The outer sphere (white) is considered massless.
The distance between the centre of
mass $O$ and the centre of symmetry $C$ is $a\ge0$,
and the angle between the principal
axis corresponding to $\IC$ and the line joining the centres
of mass and symmetry is denoted $\bdelta$ (for $a=0$, the
angle $\bdelta$ is undefined). For the \RnR, 
$\bdelta=0$.}

\label{fig:LoadedSphere}.
\end{center}
\end{figure}

\begin{figure}
\begin{center}
\includegraphics[scale=0.5]{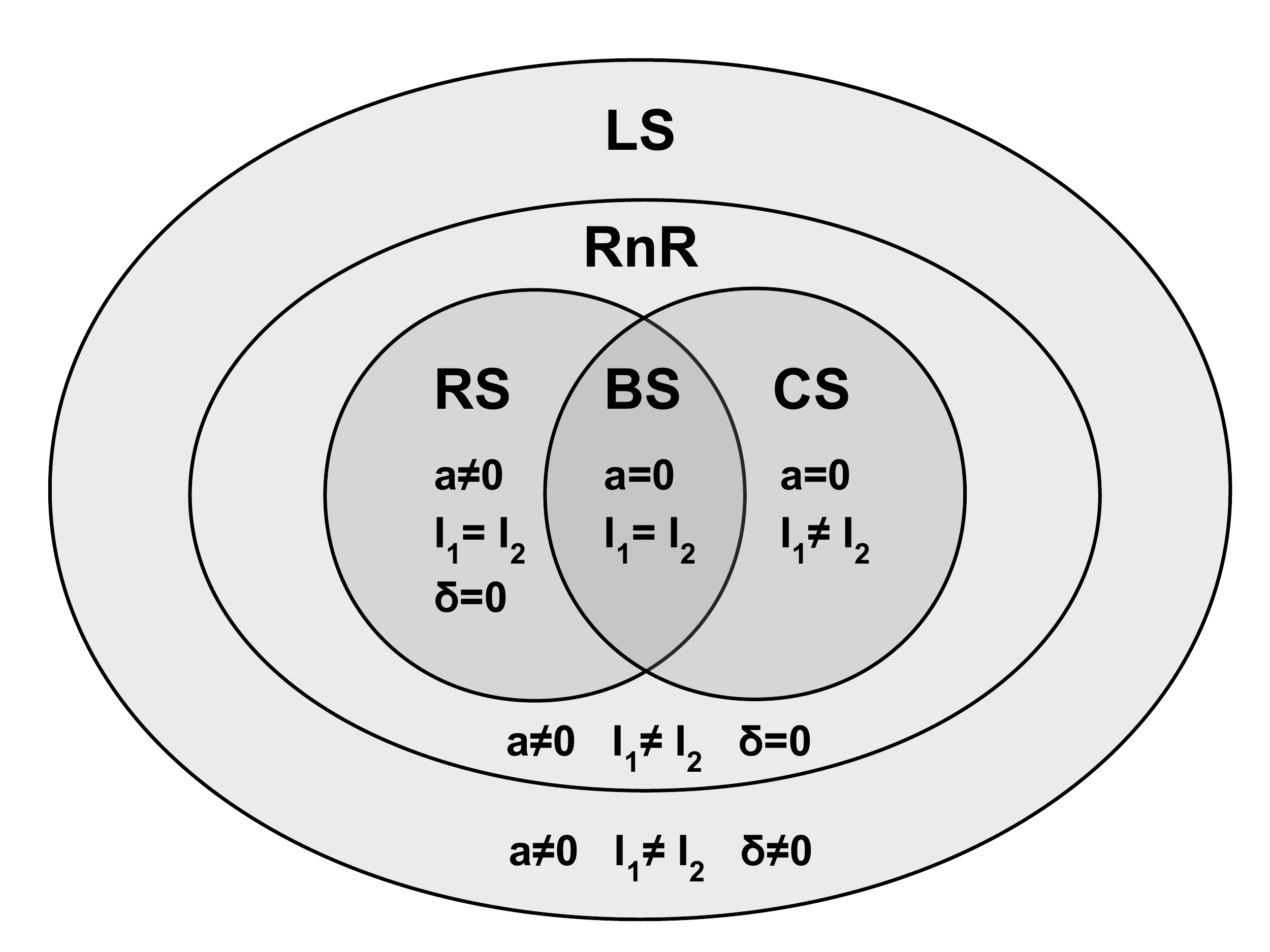}

\caption
{A hierarchy of eccentric spherical bodies. The moments of
inertia are $\IA\le\IB\le\IC$, the distance between the centres of
mass and symmetry is $a\ge0$, and the angle between the principal
axis corresponding to $\IC$ and the line joining the centres
of mass and symmetry is denoted $\bdelta$ (for $a=0$,
$\bdelta$ is undefined).
\newline
{\bf LS:} Loaded sphere: $\IA<\IB<\IC$, $a>0$, centre of sphere
{\em not} on a principal axis ($\bdelta\ne0$).
{\bf RnR:} \RnR: $\IA<\IB<\IC$, $a>0$,
principal axis through centre of sphere ($\bdelta=0$).
{\bf RS:} Routh Sphere: $\IA=\IB<\IC$, $a>0$, $\bdelta=0$.
{\bf CS:} Chaplygin Sphere: $\IA<\IB<\IC$, $a=0$.
{\bf BS:} Bobylev Sphere: $\IA=\IB<\IC$, $a=0$. }
\label{fig:ChaplyginSpheres}.
\end{center}
\end{figure}

The dynamics of a variety of rolling spherical bodies
with non-uniform distribution of mass have been studied
extensively for more than a century.
We refer to such bodies as \emph{loaded spheres}.
We can realize a loaded sphere as a massive triaxial
ellipsoid embedded eccentrically in a massless sphere
(Fig.~\ref{fig:LoadedSphere}).
The three moments of inertia about the centre of mass are
$\IA\le\IB\le\IC$, the distance between the centres of
mass and symmetry is $a\ge0$, and the angle between the principal
axis corresponding to $\IC$ and the line joining the centres
of mass and symmetry is denoted $\bdelta$
(Figure~\ref{fig:ChaplyginSpheres}).
Loaded spheres were investigated by Chaplygin
\cite{Chaplygin1897}, who obtained 
solutions in a number of particular cases. These bodies
have been discussed in several recent publications
\cite{BorisovMamaev02, Cushman98, Duistermaat04, KimByungsoo11,
Kilin01, Koslov02}.
Earlier literature is reviewed by \cite{GrayNickel00} and
a modern treatment of the dynamics of the loaded sphere
is contained in \cite{Holm2011}.

If the centre of mass and the geometric centre coincide,
we call the body the Chaplygin Sphere
(CS; See Fig.~\ref{fig:ChaplyginSpheres}).
Chaplygin \cite{Chaplygin1903}
analysed this case in detail, giving a fairly complete solution.
In general, the geometric centre does not lie on an inertial axis.
When it does, we call the body the \RnR\ (RnR). If, in addition,
the two moments of inertia transverse to this axis are equal,
$\IA=\IB<\IC$, the body is called the Routh Sphere \cite{Routh05}.
The case when both these conditions are met --- centre of mass
at the centre of sphere and $\IA=\IB$ --- has been
called Bobylev's Sphere \cite{Bobylev1892, Duistermaat04}.
Clearly, Bobylev's Sphere is a special case of both the Routh
Sphere and the Chaplygin Sphere.
The relationship between the various
bodies is shown in Fig.~\ref{fig:ChaplyginSpheres}.


\begin{figure}
\begin{center}
\includegraphics[scale=0.35]{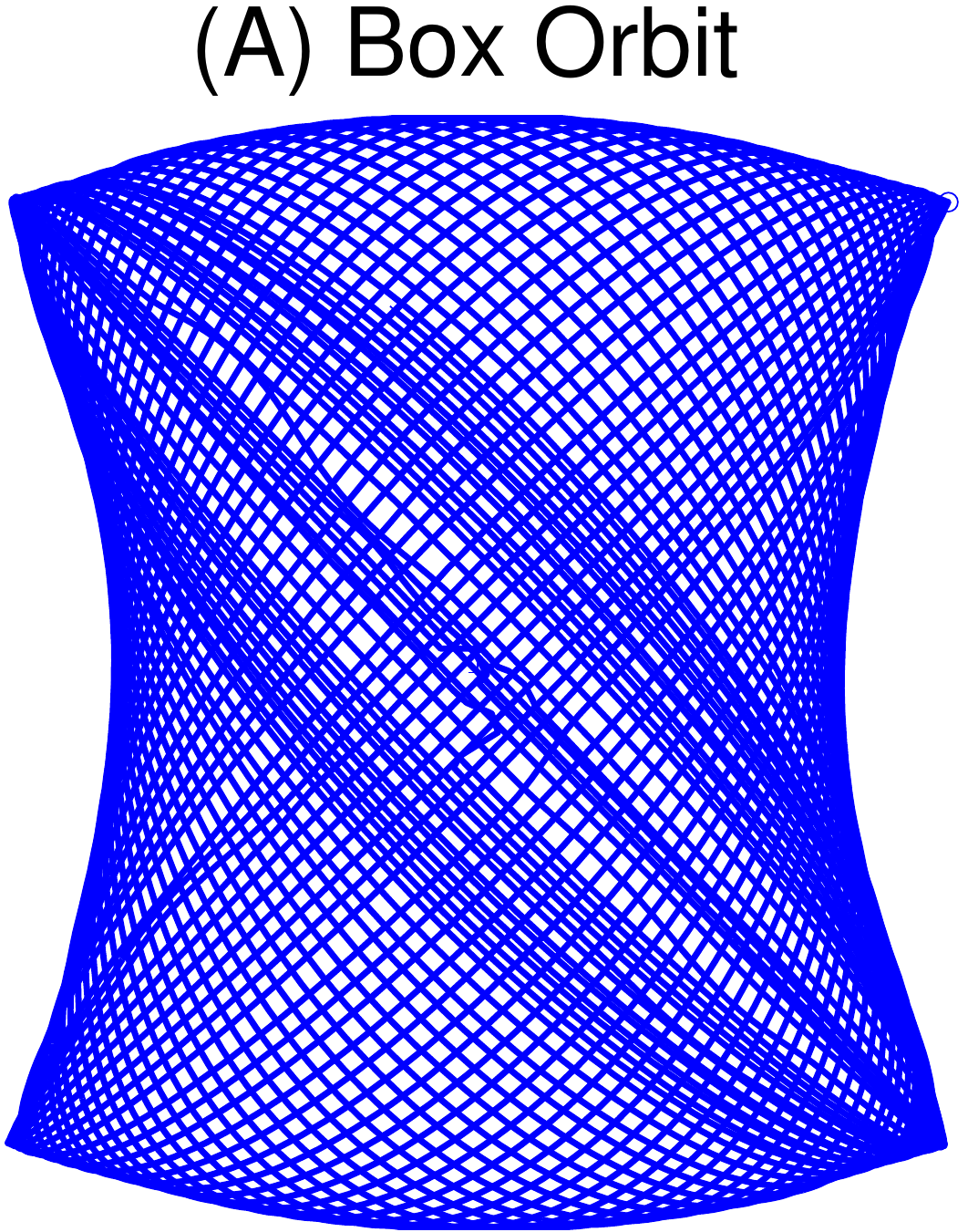}
\hspace{15mm}
\includegraphics[scale=0.35]{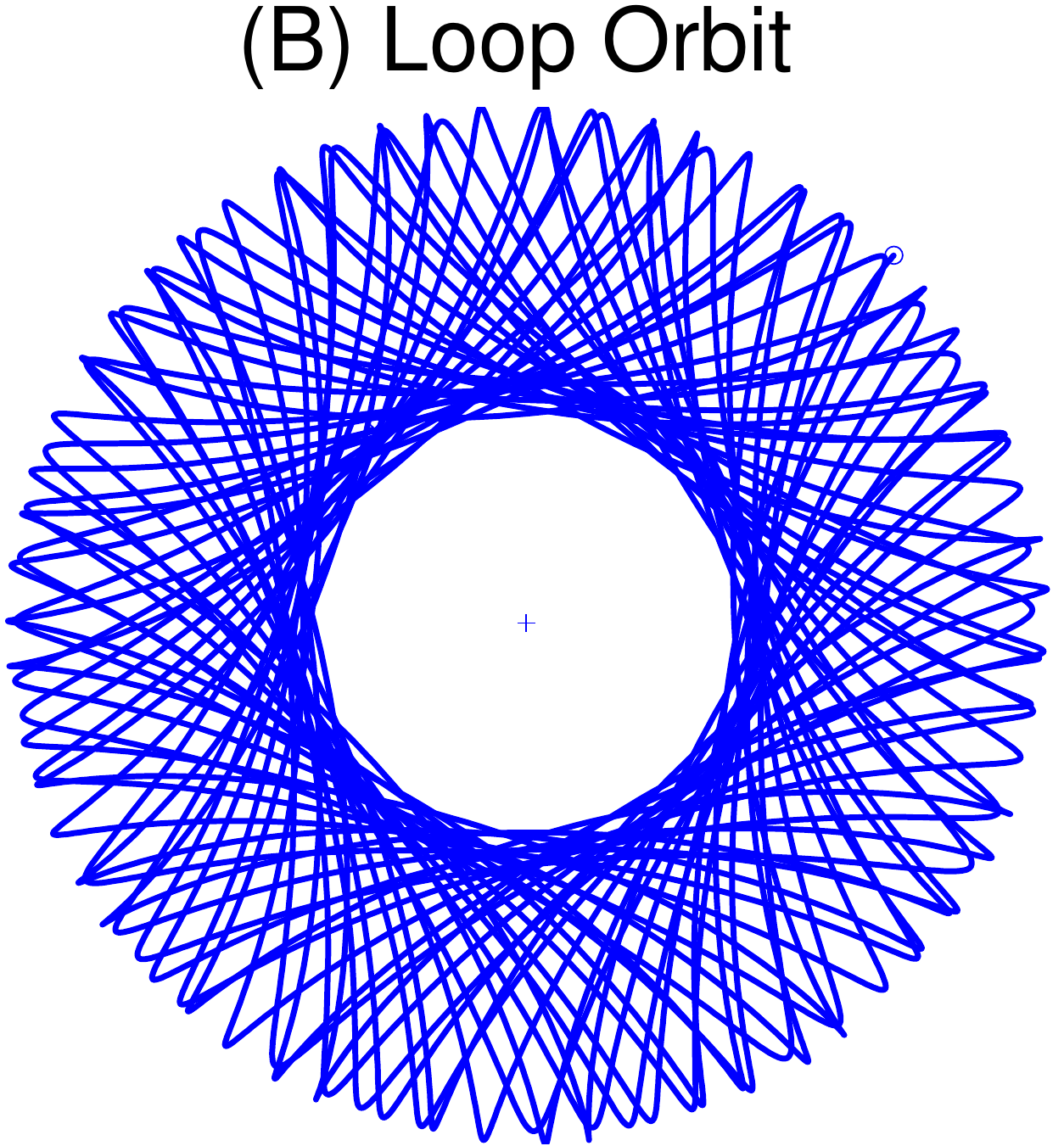}
\caption{
Projection of the trajectory of the \RnR\
in the $\theta$--$\phi$-plane ($\theta$ radial, $\phi$ azimuthal)
with $\epsilon=0.1$ for two solutions differing only in their
initial conditions.
Panel A: $\theta(0)=\pi/16$, $\phi(0)=\pi/4$, $\psi(0)=\pi/8$ and
$\omega_1(0)=\omega_2(0)=\omega_3(0)=0$.
Panel B: all parameters as before except $\omega_3(0)=0.5$}
\label{fig:RnR1}
\end{center}
\end{figure}

The dynamics of the \RnR\ were considered in \cite{L&B09}.
The orientation of the body is given by the Euler angles $\phi$,
$\theta$, and $\psi$.
In the case of the Routh Sphere ($I_1=I_2<I_3$),
there are two simple motions: pure rocking in which $\theta$
varies periodically with $\phi$ and $\psi$ constant (mod~$\pi$);
and pure rolling with $\theta$ constant and $\phi$ and $\psi$
varying steadily. 
The general motion combines these two modes of oscillation.
The azimuthal angle $\phi$ at which the polar angle $\theta$
takes its maximum values increases or decreases regularly and
monotonically.  This process is called \emph{precession}. 
When the symmetry $I_1=I_2$ is broken,
we get the \RnR, and there is a wider range of possible motions.
%
%
The direction of precession changes intermittently. 
In \cite{L&B09} we used the term \emph{recession} to describe
this reversal of precession. For a given energy level,
recession may or may not occur, depending upon the initial conditions.
In Fig.~\ref{fig:RnR1} we show the projection of the trajectory of
the \RnR\ onto the $\theta$--$\phi$-plane
($\theta$ radial, $\phi$ azimuthal in plot)
for two solutions differing only in their initial conditions.
In the left panel, the direction of precession reverses periodically.
Clearly, the angular momentum about the vertical takes both
positive and negative values.  This trajectory has recession, and is an
example of a box orbit. 
In the right panel, the trajectory circulates always in one direction
about the centre. While the angular momentum
about the vertical is not constant, it is of constant sign.
There is no recession; this is an example of a loop orbit.

\section{Perturbed Simple Harmonic Oscillator}
\label{sec:PSHO}

Many dynamical features of complex physical systems
are exhibited clearly in a very simple system, the 
perturbed simple harmonic oscillator (SHO).
The unperturbed system is the two-dimensional SHO with equal
frequencies, having the Lagrangian
\[
L_0 = \half(\dot x^2+\dot y^2) - \half\omega_0^2(x^2+y^2) 
    = \half(\dot r^2+r^2\dot\theta^2) - \half\omega_0^2 r^2 \,,
\]
where the notation is conventional.
%
%
The generic solution of this system represents motion in an
ellipse centered on the origin.
This analytical solution will serve as the basis
of a perturbation analysis. The full system that we will
study has a Lagrangian
\[
L = L_0 - \delta y^2 - \epsilon r^4 \,,
\]
where $\delta\ll \omega_0^2$ and $\epsilon\ll 1$.  The $\delta$-term
represents a breaking of the $1:1$ resonance of the system $L_0$.
The $\epsilon$-term represents a radially symmetric
stiffening of the restoring force, which results in nonlinear equations
of motion.

For $\delta>0$ and $\epsilon=0$, the Hamiltonian is separable:
\[
H = H_x + H_y = 
\half[\dot x^2+\omega_0^2 x^2] 
+ \half[\dot y^2+(\omega_0^2+2\delta) y^2] \,,
\]
and both components are constant. 
An analytical solution is immediately found:
\begin{eqnarray}
x &=& x_0 \cos \omega_0(t-t_1)             \label{eq:pertSHMx} \\
y &=& y_0 \cos(1+\delta^\prime)\omega_0(t-t_2)     \label{eq:pertSHMy}
\end{eqnarray}
where
$\delta^\prime=\sqrt{1+2\delta/\omega_0^2}-1\approx\delta/\omega_0^2$.
The generic orbit densely fills a rectangular region in the $x$--$y$-plane.
If $\delta^\prime$ is rational, the orbit is a
Lissajous figure and the motion is periodic.

The time evolution of the angular momentum $J= x\dot y-y\dot x$
is described by
\[
\dot J = -2\delta xy
\]
and clearly $J$ is not conserved, taking both positive and
negative values.

For $\delta=0$ and $\epsilon>0$, the restoring force is central and the angular 
momentum $J$ is conserved. The equation for the radial component is
\[
\ddot r + \omega_0^2 r + 4\epsilon r^3 -\displaystyle{\frac{2J^2}{r^3}} = 0
\]
and an analytical solution for $r$ in terms of elliptic integrals may easily be found.
Since the force is central, the angular momentum is conserved.
The azimuthal angle $\theta$ follows from integrating the expression for the constant
angular momentum, $J=r^2\dot\theta$. 

\begin{figure}
\begin{center}
\includegraphics[scale=0.50]{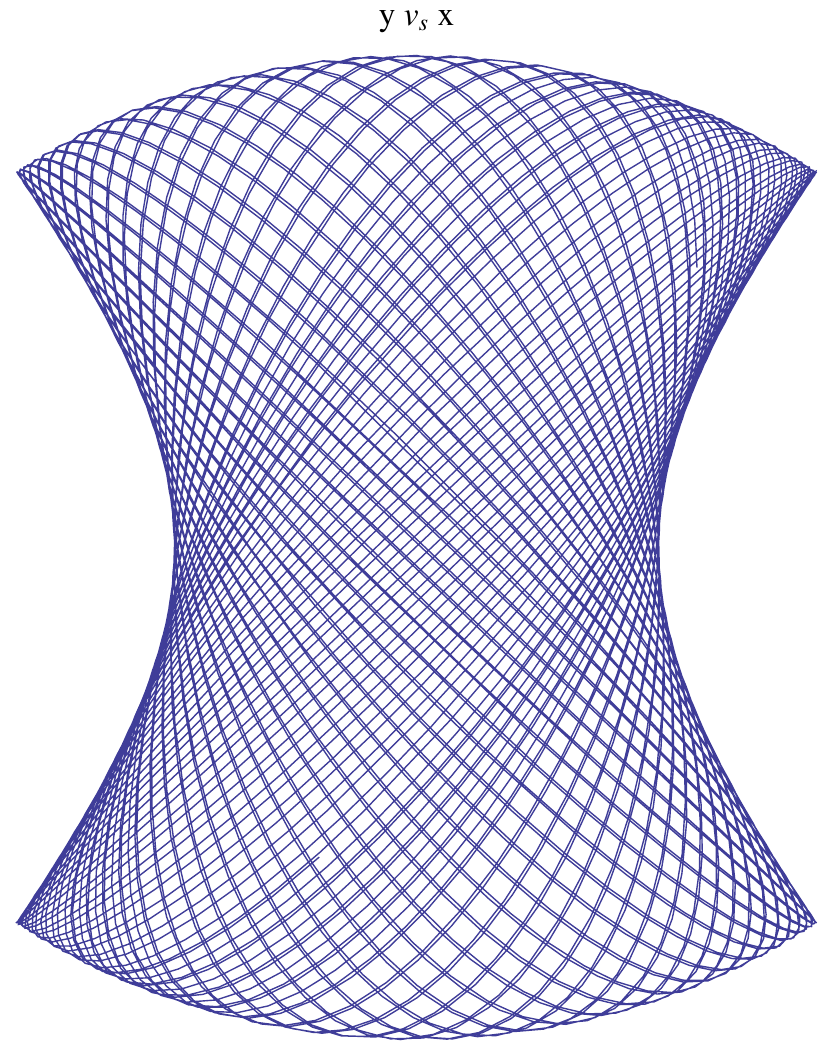}
\hspace{1cm}
\includegraphics[scale=0.50]{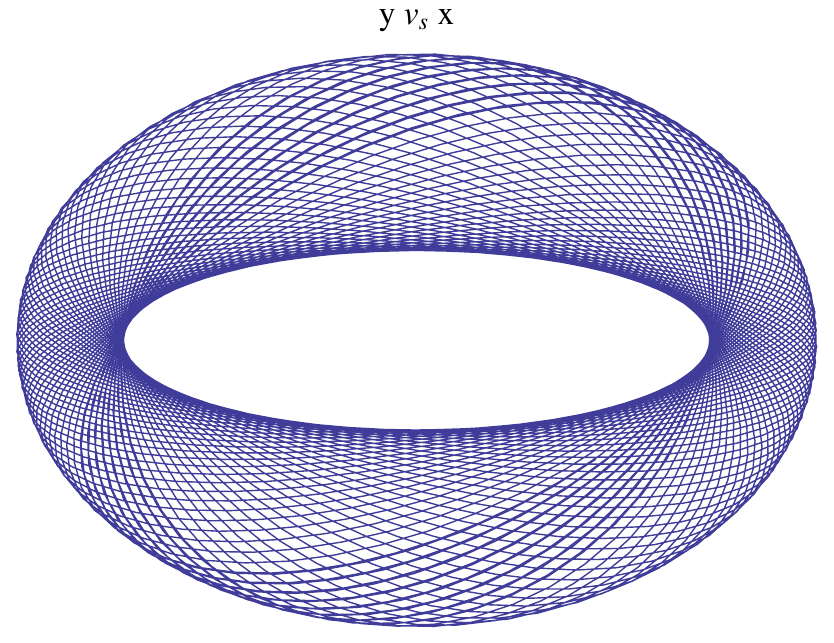}
\caption{Box and Loop orbits for the perturbed SHO. In both cases,
$\lambda = 2\epsilon U/\delta = 2.236$. 
All other parameters are equal except the
initial position and velocity. Left panel: 
$x(0)\propto\cos(0.05)$, $y(0)\propto\sin(0.05)$. 
Right panel:
$x(0)\propto\cos(0.5)$, $y(0)\propto\sin(0.5)$.}
\label{fig:BoxLoop}
\end{center}
\end{figure}

When both $\delta$ and $\epsilon$ are non-zero, an analytical
solution is not so easily found, but numerical integrations produce solutions
of both the box orbit and loop orbit types (Fig.~\ref{fig:BoxLoop}).
To analyse the system, we apply the average Lagrangian technique \cite{Whitham74}.
The solution is assumed to be of the form
\[
x(t) = \Re\{A(t)\exp(i\omega_0 t)\} \qquad
y(t) = \Re\{B(t)\exp(i\omega_0 t)\} \qquad
\]
where the amplitudes $A(t)$ and $B(t)$ are assumed to be
slowly varying compared to the exponential terms. Averaging
over the period $2\pi/\omega_0$ of the fast motion we get
\begin{eqnarray*}
\langle L \rangle &=& \frak{1}{4}\Big[
i\omega_0(A \dot{\bar{A}}-\dot A {\bar{A}}+B \dot{\bar{B}}-\dot B{\bar{B}}) 
-\delta B\bar B \\
&& \quad -\epsilon
(\frak{3}{2}|A|^4+2|A|^2|B|^2+\frak{3}{2}|B|^4+\Re\{A\bar B\}^2)
\Big]
\end{eqnarray*}
(overbars denote complex conjugates).
Introducing the modulus and phase of $A$ and $B$
\[
A = |A|\exp(i\alpha) \qquad B = |B|\exp(i\beta) 
\]
the average Lagrangian becomes
\begin{eqnarray*}
\langle L \rangle &=&
\half\omega_0(|A|^2\dot\alpha+|B|^2\dot\beta) \\
& & -\quarter\delta|B|^2 -{\textstyle\frac{1}{8}}\epsilon
[3|A|^4+4|A|^2|B|^2+3|B|^4+2|A|^2|B|^2\cos(\alpha-\beta)]
\end{eqnarray*}
We define new coordinates:
\[
U = |A|^2+|B|^2 \,, \quad V = |A|^2-|B|^2 \,, \quad
\psi=\alpha+\beta\,, \quad \phi=\alpha-\beta
\]
Then the Euler-Lagrange equations imply that $U$ is a constant of the motion. 
It represents to lowest order the value of the total energy
\[
E_0 = 
\langle\, \half(\dot x^2+\dot y^2) + \half\omega_0^2(x^2+y^2)\, \rangle
=\half\omega_0^2 U \,.
\]
Constancy of $U$ also follows, as the angle $\psi$ is an ignorable
coordinate ($\langle L \rangle$ is independent of $\psi$).
The equations for $V$ and $\phi$ are
\begin{eqnarray}
\frac{dV}{dt} &=& \left(\frac{1}{2\omega_0}\right)\epsilon(U^2-V^2)\sin 2\phi 
\label{eq:Vdot}
\\
\frac{d\phi}{dt} &=& \left(\frac{1}{2\omega_0}\right)[\epsilon(1-\cos
2\phi)V-\delta]
\label{eq:phidot}
\end{eqnarray}

We can derive an equation for the average angular momentum,
\[
\langle J \rangle = \omega_0\Im\{A\bar B\} = \half\omega_0\sqrt{U^2-V^2}\sin\phi
\,.
\]
This shows that $\langle J \rangle$ will change sign if $\phi$ passes through
any of the values $n\pi$. We will see that for box orbits $\phi$ is unbounded
whereas for loop orbits it is confined to an interval within $(n\pi,(n+1)\pi)$
so that the angular momentum does not change sign.

Defining a new variable $W=V/U$, whose physical range is the interval
$[-1,1]$, and re-scaling time by $\tau=(\delta/2\omega_0)t$,
(\ref{eq:Vdot}) and (\ref{eq:phidot}) become
\begin{eqnarray}
\frac{dW}{d\tau} &=& \lambda(1-W^2)\sin\phi \cos\phi
\label{eq:Wdot}
\\
\frac{d\phi}{d\tau} &=& \lambda W\sin^2\phi - 1
\label{eq:phidotbis}
\end{eqnarray}
where $\lambda=2\epsilon U/\delta$ is a non-dimensional parameter.
It is straightforward to show that (\ref{eq:Wdot}) and (\ref{eq:phidotbis})
are the canonical equations arising from the Hamiltonian
\[
h = \half\lambda(1-W^2)\sin^2\phi + W  \,.
\]

\begin{figure}
\begin{center}
\includegraphics[scale=0.50]{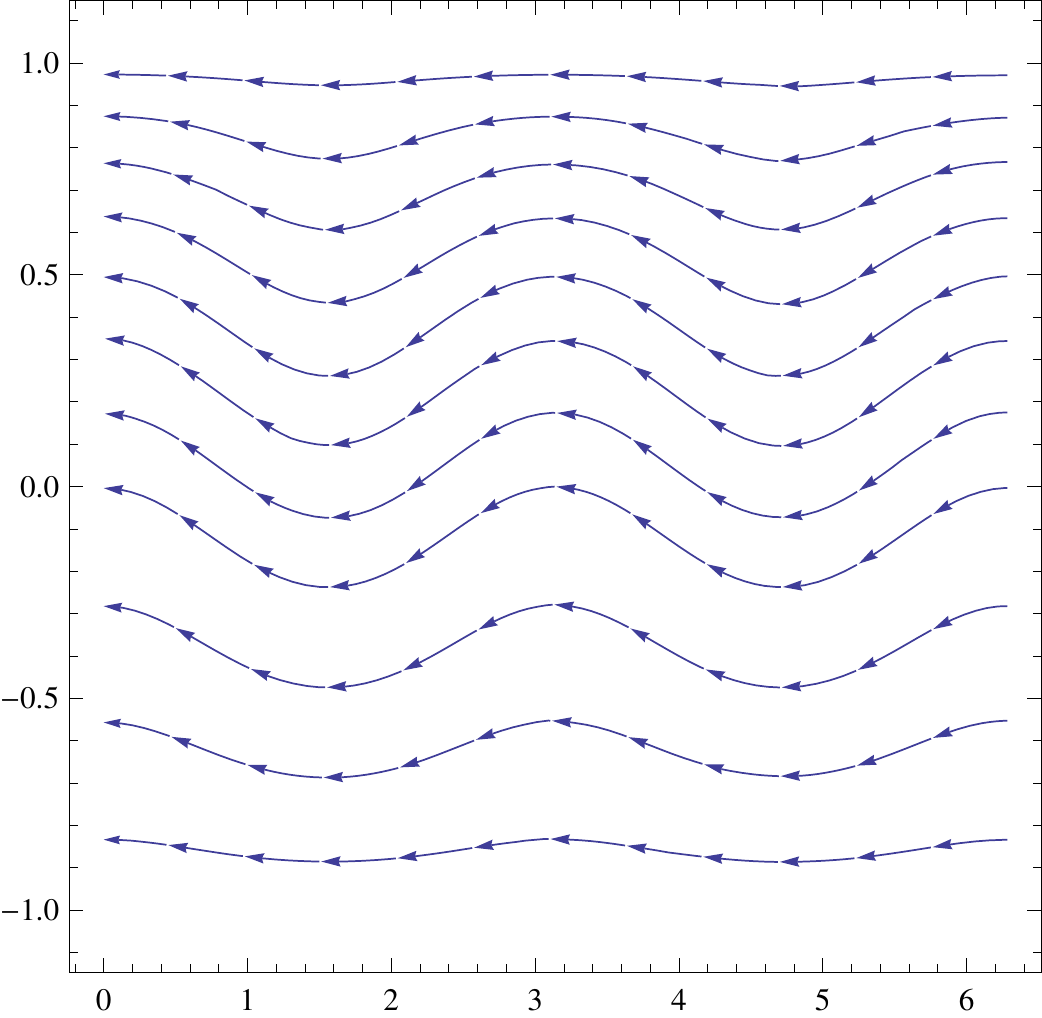}
\hspace{1cm}
\includegraphics[scale=0.50]{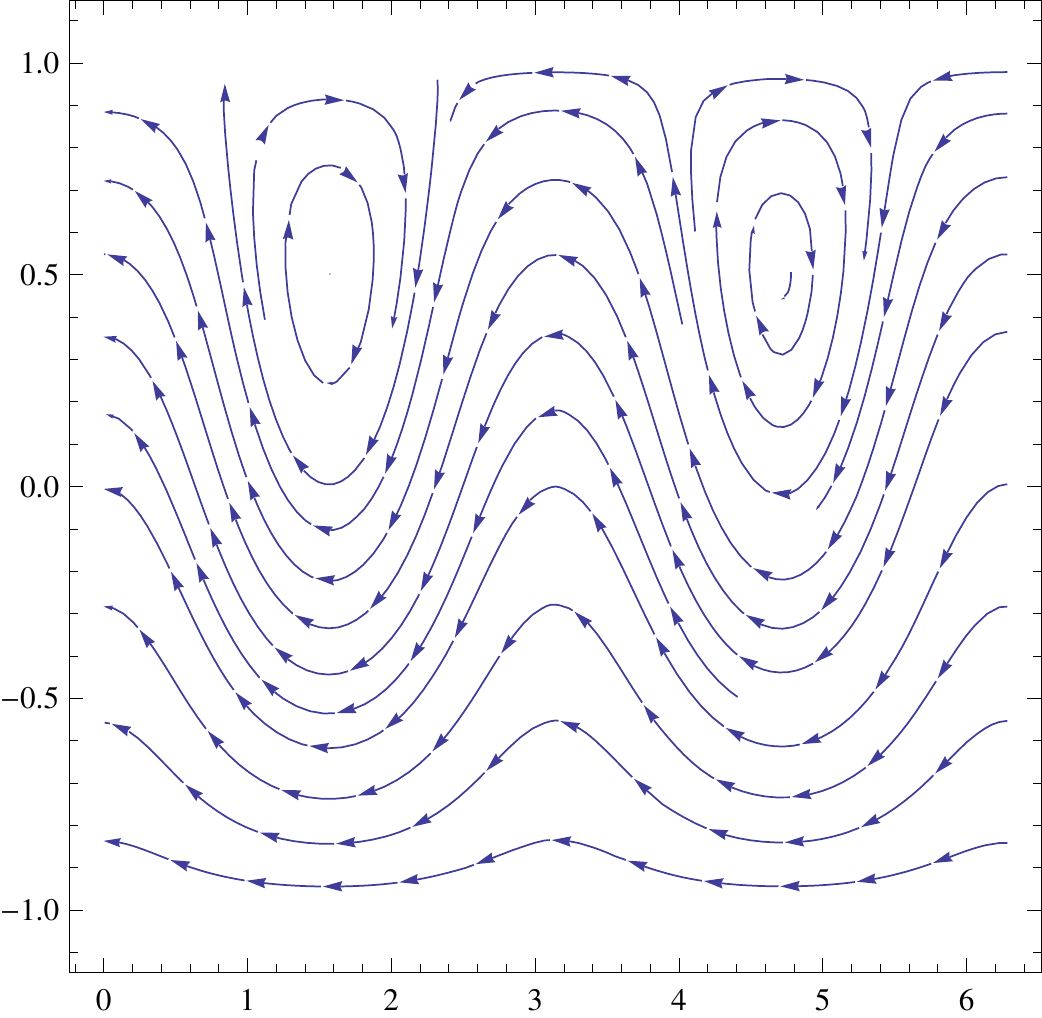}
\caption
{Phase portraits of the perturbed SHO. Left panel: $\lambda=0.5$
(only box orbits exist).
Right panel: $\lambda=2.0$ (both box and loop orbits exist).
Horizontal axis $\phi$, range
$[0,2\pi]$; vertical axis $W$, range $[-1,+1]$.}
\label{fig:PhasePortraits}
\end{center}
\end{figure}

Equations (\ref{eq:Wdot}) and (\ref{eq:phidotbis}) have equilibrium
points when $d\phi/d\tau=dW/d\tau=0$. For $\lambda<1$ there are no such points.
For $\lambda>1$ there are equilibrium points at 
$(\phi,W)=(\pi/2,1/\lambda)$ and $(\phi,W)=(3\pi/2,1/\lambda)$
which are elliptic points or centres.
There are also four equilibrium points on the boundary $W=1$, where
$\sin^2\phi=|1/\lambda|$. These are hyperbolic or saddle points.
The phase portraits in the $\phi$--$W$-plane are shown in
Fig.~\ref{fig:PhasePortraits}
(left panel: $\lambda=0.5$; right panel: $\lambda=2.0$).
%
%
The two saddle points with 
$(\sin\phi,W)=(+1/\sqrt{\lambda},1)$ are joined by heteroclinic orbits,
as are the two points with $(\sin\phi,W)=(-1/\sqrt{\lambda},1)$. 
These heteroclinic orbits separate the possible motions into two species.
Below (or outside) the separatrices,
the variable $\phi$ decreases continually,
and the angular momentum 
$\langle J \rangle = \half\omega_0\sqrt{U^2-V^2}\sin\phi$ changes sign
periodically. These trajectories correspond to box orbits.
Above (or within) the separatrices,
the trajectories surround the centres and the variable $\phi$ is
confined to an interval within $(n\pi,(n+1)\pi)$.
Thus the angular momentum is of a
single sign.  Such trajectories correspond to loop orbits.

The value $\lambda=1$ is a bifurcation point for the system
(\ref{eq:Wdot})--(\ref{eq:phidotbis}).
The line $W=1$ corresponds to $B\equiv 0$ and represents an oscillation along
the $x$-axis.
Likewise, the line $W=-1$ corresponds to $A\equiv 0$ and represents an oscillation
along the $y$-axis. For $\lambda<1$ both these motions are stable. For
$\lambda>1$ the former is unstable while the latter is stable.


\section{Equations of Motion of the \RnR}
\label{sec:RnR-Eqns}

The \RnR\ and its symmetric counterpart, the Routh Sphere,
were briefly described in \S\ref{sec:Rocknroller}.
Here we present the general equations for the motion of
the \RnR\ and also the simplified equations for small amplitude motions.

\subsection{Euler angle equations for finite amplitude motion}
\label{sec:EulerAngleEqns}

The equations of motion in terms of Euler angles are given in
\cite{L&B09}:
\begin{equation}
\bSigma\bthdot = {\bomega} \,, \qquad \bi{K}\bomdot
               = \bi{P}_{\bomega} \label{eq:thetaomega}
\label{eq:BasicEqns}
\end{equation}
where
$$\fl \bthdot = \left( \begin{matrix}
 { \dot\theta \cr \dot\phi \cr \dot\psi }
\end{matrix}  \right) \,, \qquad
\bomdot = \left( \begin{matrix}
 { \dot\omega_1 \cr \dot\omega_2 \cr \dot\omega_3 }
\end{matrix}  \right) \,,
$$
the matrices $\bSigma$ and $\bi{K}$ are
$$
\bSigma = \left[ \begin{matrix}
  { \chi   &  s\sigma  &  0  \cr
  -\sigma  &   s\chi   &  0  \cr
     0     &     c     &  1   }
\end{matrix}  \right]
\qquad \bi{K} = \left[ \matrix{
        \IA+f^2+s^2\chi^2 & -s^2\sigma\chi      & -fs\sigma \cr
         -s^2\sigma\chi   & \IB+f^2+s^2\sigma^2 & -fs\chi   \cr
            -fs\sigma     & -fs\chi             &  \IC+s^2  }
\right]
\nonumber
$$
and the vector $\bi{P}_{\bomega}$ is
\[
\bi{P}_{\bomega} = \left(
\begin{array}{l}
       -(g+\omega_1^2+\omega_2^2)as\chi
       +(\IB-\IC-af)\omega_2\omega_3  \cr
\phantom{+}(g+\omega_1^2+\omega_2^2)as\sigma
       +(\IC-\IA+af)\omega_1 \omega_3 \cr
\phantom{+}(\IA-\IB)\omega_1\omega_2
       + as( -\chi\omega_1+\sigma\omega_2)\omega_3
\end{array}
\right)\,,
\]
where $s=\sin\theta$, $c=\cos\theta$, $f=c-a$, $\chi=\cos\psi$ and $\sigma=\sin\psi$.
Unit mass and radius are assumed and $a$ is the distance from the geometric
centre to the centre of mass. 
Note that neither $\bi{K}$ nor $\bi{\bi{P}_{\bomega}}$ depends explicitly on
$\phi$. Thus $\phi$ is an ignorable coordinate in the system
(\ref{eq:thetaomega}).

To study the fundamental oscillations of the system, we consider motions near
the stable equilibrium point $\theta=0$. An attempt to linearize
(\ref{eq:BasicEqns}) directly, assuming $\theta$ to be small, do not lead to
a tractable system: the singularities of $\dot\phi$ and $\dot\psi$ when
$\theta=0$ thwart our endeavours. Thus we are led to seek a system
of coordinates that circumvents these singularities and yields simple linear
equations for small $\theta$. The unit quaternions provide a suitable system.


\subsection{Quaternionic equations for small amplitude motion}
\label{sec:Quaternion-Eqns}


The Euler angles relate the orientation of the body frame to
that of the space frame. This relationship may also be expressed
in terms of Euler's symmetric parameters, or the
Euler-Rodrigues parameters \cite{Altmann86, Whittaker37}, 
defined by
\begin{eqnarray*}
\gamma = \cos\half\theta\cos\half(\phi+\psi)
\quad&&\quad
\xi    = \sin\half\theta\cos\half(\phi-\psi)
\\
\zeta  = \cos\half\theta\sin\half(\phi+\psi)
\quad&&\quad
\eta   = \sin\half\theta\sin\half(\phi-\psi) 
\end{eqnarray*}
(the notation here differs slightly from
\cite{Whittaker37}; see Appendix A).
These parameters satisfy the relationship
$\gamma^2+\zeta^2+\xi^2+\eta^2=1$.
They are the components of a unit quaternion
$\mathbf{\mathsf{q}} = \gamma + \xi{\bf i} + \eta{\bf j} + \zeta{\bf k}$.
 
Expressions for the angular rates of change follow in a straightforward manner:
\begin{eqnarray*}
\dot\theta &=& 
\frac{(\xi\dot\xi+\eta\dot\eta)-(\gamma\dot\gamma+\zeta\dot\zeta)}
     {\sqrt{(\xi^2+\eta^2)(\gamma^2+\zeta^2)}} \\
\dot\phi &=& 
\left(\frac{\gamma\dot\zeta-\zeta\dot\gamma}{\gamma^2+\zeta^2}\right) + 
\left(\frac{\xi\dot\eta-\eta\dot\xi}{\xi^2+\eta^2}\right)  \\
\dot\psi &=& 
\left(\frac{\gamma\dot\zeta-\zeta\dot\gamma}{\gamma^2+\zeta^2}\right) -
\left(\frac{\xi\dot\eta-\eta\dot\xi}{\xi^2+\eta^2}\right)  
\end{eqnarray*}
Moreover,
\begin{eqnarray*}
s &= 2\sqrt{(\gamma^2+\zeta^2)(\xi^2+\eta^2)}
\qquad\qquad
c &= (\gamma^2+\zeta^2)-(\xi^2+\eta^2) 
\\
\chi &= \frac{\gamma\xi+\zeta\eta}{\sqrt{(\gamma^2+\zeta^2)(\xi^2+\eta^2)}}
\qquad\qquad
\sigma &= \frac{\zeta\xi-\gamma\eta}{\sqrt{(\gamma^2+\zeta^2)(\xi^2+\eta^2)}}
\\
s_\phi &= \frac{\gamma\eta+\zeta\xi}{\sqrt{(\gamma^2+\zeta^2)(\xi^2+\eta^2)}} 
\qquad\qquad
c_\phi &= \frac{\gamma\xi-\zeta\eta}{\sqrt{(\gamma^2+\zeta^2)(\xi^2+\eta^2)}} 
\end{eqnarray*}
where $s_\phi=\sin\phi$ and $c_\phi=\cos\phi$.
The components of angular velocity are
\begin{eqnarray}
\omega_1 &=& 2[\gamma\dot\xi-\xi\dot\gamma+\zeta\dot\eta-\eta\dot\zeta]
\nonumber \\
\omega_2 &=& 2[\gamma\dot\eta-\eta\dot\gamma+\xi\dot\zeta-\zeta\dot\xi]
\label{eq:Appomega} \\
\omega_3 &=& 2[\gamma\dot\zeta-\zeta\dot\gamma+\eta\dot\xi-\xi\dot\eta]
\nonumber
\end{eqnarray}


At first order in small $\theta$ we may write the Euler-Rodrigues
parameters as
\begin{eqnarray*}
\gamma = \cos\half(\phi+\psi) =O(1)
\quad&&\quad
\xi    = \half\theta\cos\half(\phi-\psi) =O(\theta)
\\
\zeta  = \sin\half(\phi+\psi) =O(1)
\quad&&\quad
\eta   = \half\theta\sin\half(\phi-\psi)  =O(\theta)
\end{eqnarray*}
Moreover, at this order of approximation,
$$
\omega_1=O(\theta)   \qquad \omega_2=O(\theta)  \qquad
   \omega_3=2(\gamma \dot\zeta - \zeta \dot\gamma)=O(1)
$$
%
The third equation of (\ref{eq:BasicEqns}) reduces to
$\dot\omega_3=O(\theta^2)$,
so we can take $\omega_3$ to be constant.
The order-one quaternion elements, $\gamma$ and $\zeta$,
are easily found: combining
$$
\gamma \dot\gamma + \zeta \dot\zeta = 0 \quad\mbox{and}\quad 
\gamma \dot\zeta - \zeta \dot\gamma = \half\omega_3
$$
we see that $\dot\gamma=-\half\omega_3\zeta$ and
$\dot\zeta=+\half\omega_3\gamma$,
which are immediately solved to yield
\begin{equation}
\gamma = \cos \half\omega_3(t-t_{00}) \,, \qquad
\zeta  = \sin \half\omega_3(t-t_{00}) 
\label{eq:gammazeta}
\end{equation}
Here we have used $\gamma^2+\zeta^2=1$. We choose the time origin such that
$t_{00}=0$.
The remaining two equations of (\ref{eq:BasicEqns}) may now be written
\begin{eqnarray}
\gamma\ddot\xi  + \zeta \ddot\eta - \kappa_{21}\omega_3(\zeta \dot\xi  -
\gamma\dot\eta) + \Omega_1^2(\gamma\xi  + \zeta \eta) &=& 0 
\label{eq:ddotxieta1} \\
\zeta \ddot\xi  - \gamma\ddot\eta + \kappa_{12}\omega_3(\gamma\dot\xi  +
\zeta \dot\eta) + \Omega_2^2(\zeta \xi  - \gamma\eta) &=& 0 
\label{eq:ddotxieta2}
\end{eqnarray}
where the constant parameters in the coefficients are
\begin{eqnarray*}
\phantom{.}\hspace{-15mm}
\kappa_{12} = \frac{\IC-\IA+af_0}{\IB+f_0^2} & \qquad
\phantom{.}\hspace{-0mm}
\Omega_{10}^2=\frac{ga}{\IA+f_0^2} & \qquad
\phantom{.}\hspace{-0mm}
\Omega_1^2=\Omega_{10}^2+\half(\kappa_{21}+\half)\omega_3^2 
\\
\phantom{.}\hspace{-15mm}
\kappa_{21} = \frac{\IC-\IB+af_0}{\IA+f_0^2} & \qquad
\phantom{.}\hspace{-0mm}
\Omega_{20}^2=\frac{ga}{\IB+f_0^2} & \qquad
\phantom{.}\hspace{-0mm}
\Omega_2^2=\Omega_{20}^2+\half(\kappa_{12}+\half)\omega_3^2
\end{eqnarray*}
with $f_0=(1-a)$. These equations may be transformed, by a simple rotation,
to a system with constant coefficients. We define
\begin{equation}
\left( \matrix { \mu \cr \nu } \right) =
\left[
\matrix{  \gamma &  \zeta  \cr
         -\zeta  &  \gamma    } \right]
\left( \matrix { \xi \cr \eta } \right) \,.
\label{eq:munudef}
\end{equation} 
The following relationships are straightforward to derive:
\begin{eqnarray*}
\omega_1 = 2(\dot\mu-\omega_3\nu) \qquad & &
\dot\theta = 2(\mu\dot\mu+\nu\dot\nu)/\sqrt{\mu^2+\nu^2} \\
\omega_2 = 2(\dot\nu+\omega_3\mu) \qquad & &
\dot\phi   = \omega_3 + (\mu\dot\nu-\nu\dot\mu)/(\mu^2+\nu^2) \\
\omega_3 = 2(\gamma\dot\zeta-\zeta\dot\gamma) \qquad & &
\dot\psi   = \phantom{\omega_3} - (\mu\dot\nu-\nu\dot\mu)/(\mu^2+\nu^2) 
\end{eqnarray*}
Equations (\ref{eq:ddotxieta1}) and (\ref{eq:ddotxieta2}) may now be written 
\begin{eqnarray}
\ddot\mu  - 2k_2\dot\nu + \tilde\Omega_1^2 \mu &=& 0 
\label{eq:ddotmunuone} \\
\ddot\nu  + 2k_1\dot\mu + \tilde\Omega_2^2 \nu &=& 0 
\label{eq:ddotmunutwo}
\end{eqnarray}
where
\begin{eqnarray*}
k_1 &= \half(1-\kappa_{12})\omega_3 \,, \qquad
\tilde\Omega_1^2  &=  \Omega_{10}^2 + \kappa_{21}\omega_3^2
\\
k_2 &= \half(1-\kappa_{21})\omega_3 \,, \qquad
\tilde\Omega_2^2  &=  \Omega_{20}^2 + \kappa_{12}\omega_3^2 \,.
\end{eqnarray*}
If we seek a solution of (\ref{eq:ddotmunuone})--(\ref{eq:ddotmunutwo})
in the form
$$
\mu = \mu_0 \cos\beta(t-t_0)  \qquad\mbox{and}\qquad
\nu = \nu_0 \sin\beta(t-t_0) 
$$
the system may be written
\begin{equation}
\left[
\matrix{ \tilde\Omega_1^2-\beta^2   & -2k_2\beta  \cr
                -2k_1\beta          &  \tilde\Omega_2^2-\beta^2   } \right]
\left( \matrix { \mu_0 \cr \nu_0 } \right)
= \left( \matrix { 0 \cr 0 } \right)
\label{eq:eeeqn}
\end{equation}
The determinant is a biquadratic in $\beta$ with four real roots, occurring in
positive and negative pairs. We denote the positive eigenvalues by $\beta_1$ and
$\beta_2$ and assume that $0\le\beta_1\le\beta_2$.  The eigenvectors are 
$(1,\lambda_1)^{\rm T}$ and $(1,\lambda_2)^{\rm T}$, with 
\begin{equation}
\lambda_1 = \frac{\tilde\Omega_1^2-\beta_1^2}{2k_2\beta_1}
          = \frac{2k_1\beta_1}{\tilde\Omega_2^2-\beta_1^2} \,,
\qquad
\lambda_2 = \frac{\tilde\Omega_1^2-\beta_2^2}{2k_2\beta_2}
          = \frac{2k_1\beta_2}{\tilde\Omega_2^2-\beta_2^2}
\label{ew:lambda12}
\end{equation}
and we can write the general solution of
(\ref{eq:ddotmunuone})--(\ref{eq:ddotmunutwo}) as
\begin{eqnarray}
\mu &= \phantom{\lambda_1}\mu_1\cos\beta_1(t-t_1)
    &+ \phantom{\lambda_2}\mu_2 \cos\beta_2(t-t_2) 
\label{eq:gensocone} \\
\nu &=          \lambda_1 \mu_1\sin\beta_1(t-t_1) 
    &+          \lambda_2 \mu_2\sin\beta_2(t-t_2)
\label{eq:gensoctwo}
\end{eqnarray}

The equations (\ref{eq:ddotmunuone})--(\ref{eq:ddotmunutwo})
are now completely solved.  The solution
(\ref{eq:gensocone})--(\ref{eq:gensoctwo}) is determined by the initial
conditions $\{\mu_1,\mu_2,t_1,t_2\}$. These are
equivalent to conditions $\{\mu(0),\dot\mu(0),\nu(0),\dot\nu(0)\}$.
Solutions of (\ref{eq:ddotxieta1})--(\ref{eq:ddotxieta2})
follow immediately by means of (\ref{eq:gammazeta}) and
(\ref{eq:munudef}).

\subsection{Lagrangian and Hamiltonian}
\label{sec:LagHam}

Equations (\ref{eq:ddotmunuone})--(\ref{eq:ddotmunutwo})
may be derived from the Lagrangian
\begin{equation}
L = \half(k_1\dot\mu^2+k_2\dot\nu^2)
  - \half(k_1\tilde\Omega_1^2\mu^2+k_2\tilde\Omega_2^2\nu^2)
  + k_1k_2(\mu\dot\nu-\nu\dot\mu)
\nonumber
\end{equation}
The generalized momenta are $p_\mu = k_1(\dot\mu-k_2\nu)$ and  
$p_\nu = k_2(\dot\nu+k_2\mu)$ and the Hamiltonian, obtained from the
Legendre transformation, is  
\begin{eqnarray}
H = \half\displaystyle{\left(\frac{p_\mu^2}{k_1}+\frac{p_\nu^2}{k_2}\right)}
   &-& [k_1\mu p_\nu-k_2\nu p_\mu] \\
   &+& \half[k_1(k_1k_2+\tilde\Omega_1^2)\mu^2+k_2(k_1k_2+\tilde\Omega_2^2)\nu^2]
\nonumber
\end{eqnarray}
The numerical value of the Hamiltonian is equal to the (constant) energy
\[
E_{\mu+\nu} = \half(k_1\dot\mu^2+k_2\dot\nu^2)
  + \half(k_1\tilde\Omega_1^2\mu^2+k_2\tilde\Omega_2^2\nu^2)
\]

An additional constant of the motion can be found from the solutions
(\ref{eq:gensocone})--(\ref{eq:gensoctwo}) for $\mu$ and $\nu$ and their
time derivatives for $\dot\mu$ and $\dot\nu$. We can solve the four expressions
for the sines and cosines in terms of $\{\mu,\nu,\dot\mu,\dot\nu\}$. These
can then be combined to yield the following constants:
\begin{eqnarray}
K_1 &\equiv 
\left(
\frac{\lambda_2\dot{\mu}+\beta_2\nu}{\beta_1\lambda_2-\beta_2\lambda_1}
\right)^2
+
\left(
\frac{\dot{\nu}-\beta_2\lambda_2\mu}{\beta_1\lambda_1-\beta_2\lambda_2}
\right)^2
&=
\mu_1^2 \,,
\\
K_2 &\equiv
\left(
\frac{\lambda_1\dot{\mu}+\beta_1\nu}{\beta_1\lambda_2-\beta_2\lambda_1}
\right)^2
+
\left(
\frac{\dot{\nu}-\beta_1\lambda_1\mu}{\beta_1\lambda_1-\beta_2\lambda_2}
\right)^2
&=
\mu_2^2 \,.
\end{eqnarray}
Numerical tests confirm that $K_1$ and $K_2$ remain constant.
They may be combined linearly to form $E_{\mu+\nu}$ and an additional
independent constant.


\section{Epi-elliptic solution for small asymmetry
$\mathbf{(\epsilon=(\IB-\IA)/\IA\ll 1)}$}
\label{sec:RnRasym}

For the symmetric case, $\epsilon=0$ and the matrix in
(\ref{eq:eeeqn}) takes the simple form
$$
\left[
\matrix{ A   & -B  \cr
        -B   &  A     } \right]
$$
so that the eigenvectors are multiples of
$(1,1)^{\rm T}$ and $(1,-1)^{\rm T}$.
For $\epsilon\ne0$, this is no longer the case. 
The eigenvalues are perturbed to $\beta_1=\beta_1^0+\delta\beta_1$ and
$\beta_2=\beta_2^0+\delta\beta_2$, where
$\beta_1^0$ and $\beta_2^0$ are the values for $\epsilon=0$.
The eigenvectors are
$(1,\lambda_1)^{\rm T}$ and $(1,\lambda_2)^{\rm T}$ 
and, for small asymmetry ($\epsilon\ll 1$), we have
$\lambda_1\approx +1$ and $\lambda_2\approx -1$.

The complete solution of the \RnR\ equations for small
amplitude motion is 
\begin{eqnarray}
\gamma &= \cos \half\omega_3t  
\label{eq:RRgamma} \\
\zeta  &= \sin \half\omega_3t  
\label{eq:RRzeta} \\
\mu &= \phantom{\lambda_1}\mu_1\cos\beta_1(t-t_1) 
    &+ \phantom{\lambda_2}\mu_2\cos\beta_2(t-t_2)
\label{eq:RRmu} \\
\nu &=          \lambda_1 \mu_1\sin\beta_1(t-t_1) 
    &+          \lambda_2 \mu_2\sin\beta_2(t-t_2)
\label{eq:RRnu}
\end{eqnarray}
The projection in the $\gamma$--$\zeta$-plane, given by
(\ref{eq:RRgamma})--(\ref{eq:RRzeta}), is a rotation with frequency
$\half\omega_3$.
The solution (\ref{eq:RRmu})--(\ref{eq:RRnu}) has two components,
each having a trajectory that is elliptic
(for the Routh Sphere, $\epsilon=0$, they are both circular).
The first component is
\[
\mu = \mu_1 \cos[\beta_1(t-t_1)] \,,
\qquad
\nu = \mu_1\lambda_1 \sin[\beta_1(t-t_1)]  \,,
\]
elliptical motion with frequency
$\beta_1$ and semi-axes $\mu_1$ and $\mu_1\lambda_1$,
counterclockwise if $\lambda_1>0$ (recall that $\lambda_1=1$ in the
limiting case $\epsilon=0$).
The second component is
\[
\mu =  \mu_2 \cos[\beta_2(t-t_2)] \,,
\qquad
\nu =  \mu_2 \lambda_2 \sin[\beta_2(t-t_2)]  \,,
\]
elliptical motion with frequency
$\beta_2$ and semi-axes $\mu_2$ and $\mu_2\lambda_2$,
clockwise if $\lambda_2<0$ (and $\lambda_2=-1$ in the
limiting case $\epsilon=0$).
The overall character of the trajectory is thus determined by the
relative magnitudes and signs of the parameters
$\{\lambda_1,\lambda_2\}$ and the initial conditions
$\{\mu_1,\mu_2\}$.

\begin{figure}
\includegraphics[scale=0.55]{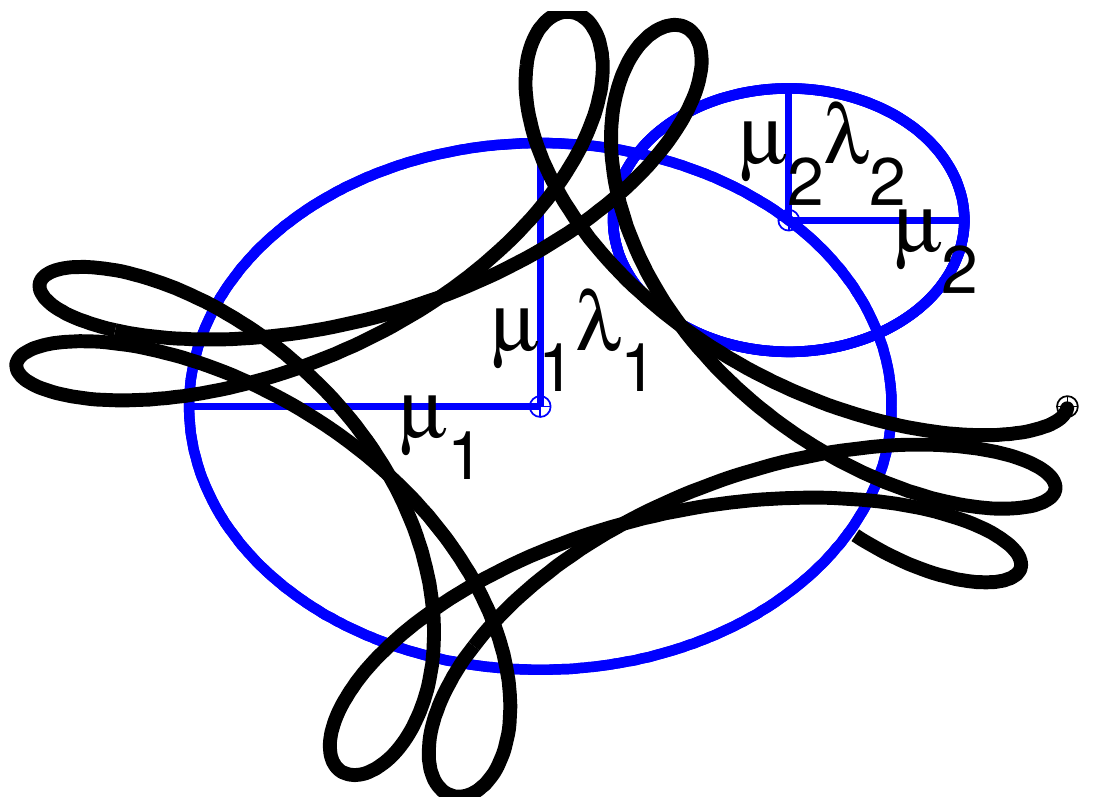}
\hspace{5mm}
\includegraphics[scale=0.55]{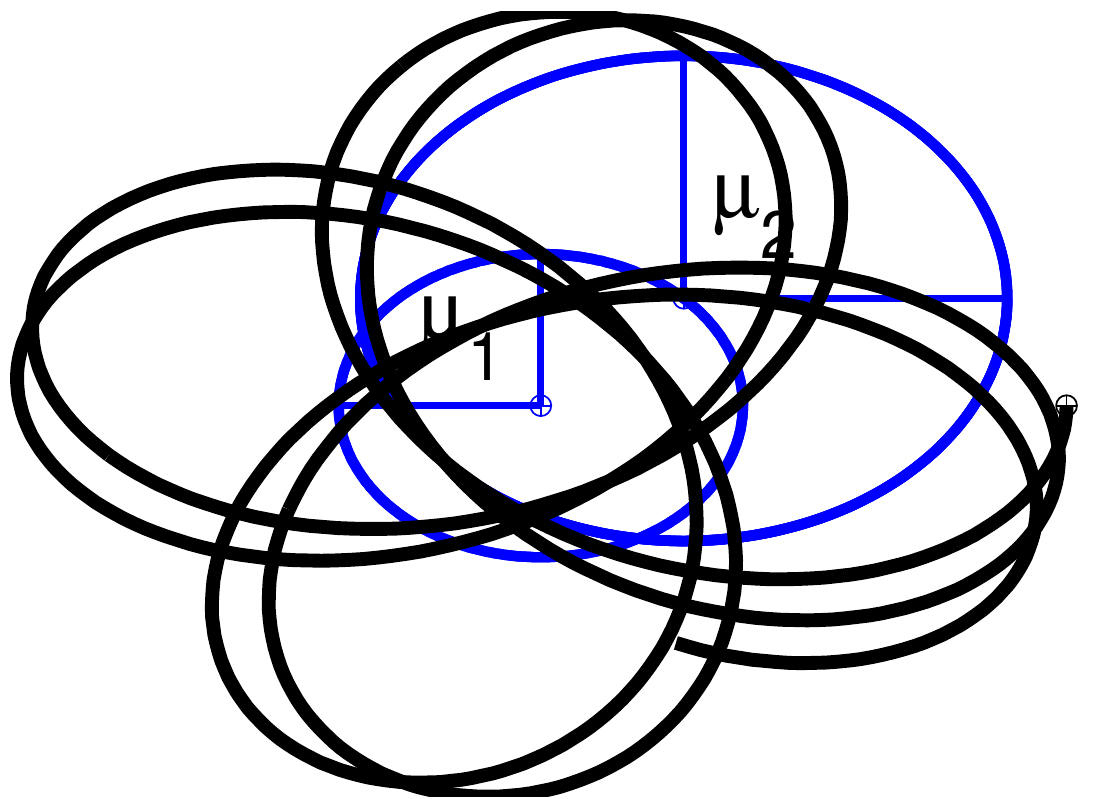}

\caption{Rock'n'roller:
the trajectories in the $\mu$--$\nu$-plane are epi-ellipses.
Left: $\mu_1=1$, $\mu_2=0.5$, $\lambda_1=0.75$, $\lambda_2=-0.75$.
Right: $\mu_1=1$, $\mu_2=1.6$, $\lambda_1=0.75$, $\lambda_2=-0.75$.
In both cases, $\beta_1=1$ and $\beta_2=1/\pi$.}
\label{fig:Epellipses}
\end{figure}

In Fig~\ref{fig:Epellipses} we illustrate two characteristic solutions
in the $\mu$--$\nu$-plane.
In both cases, $\beta_1=1/\pi$ and $\beta_2=1$.
In the left panel, $\mu_2<\mu_1$ and also $\mu_2\lambda_2<\mu_1\lambda_1$,
the orbit circulates 
about the centre, sometimes curving towards it and sometimes curving away.
We call this a \emph{centrifugal} orbit.
The representative point is bounded away from the centre.
For the solution in the right panel, $\mu_2>\mu_1$ and also
$\mu_2\lambda_2>\mu_1\lambda_1$. 
The orbit circulates about the centre, $\mu=\nu=0$,
always curving towards it. We call this a \emph{centripetal} orbit.


It is clear on geometric grounds that if $(|\mu_1|-|\mu_2|)$
and
$(|\lambda_1\mu_1|-|\lambda_2\mu_2|)$ are of the same sign,
the trajectory cannot reach the centre, $\mu=\nu=0$. Thus, the
criterion for recession may be written
\begin{equation}
(|\mu_1|-|\mu_2|)\cdot(|\lambda_1\mu_1|-|\lambda_2\mu_2|) < 0 
\label{eq:rescrit}
\end{equation}
(see Appendix~B).
In this case, the orbit may approach arbitrarily close to the
centre, and the angular momentum about the centre may change
sign.

\subsection{Special case: central orbits with $\omega_3=0$}

The case $\omega_3=0$ is especially simple:
$(\mu,\nu)=(\xi,\eta)
\approx(\half\theta\cos\phi,\half\theta\sin\phi)$, 
so the trajectories in the $\mu$--$\nu$-plane are structurally
identical to those on a polar $\theta$--$\phi$ plot
(with $\theta$ radial and $\phi$ azimuthal).
Equations (\ref{eq:ddotmunuone})--(\ref{eq:ddotmunutwo})
take a particularly simple form
\[
\ddot\mu + \Omega_{10}^2\mu
\qquad\mbox{\rm and}\qquad
\ddot\nu + \Omega_{20}^2\nu
\]
and the solution may be written immediately:
\begin{eqnarray}
\mu &=& \mu_0\cos\Omega_{10}(t-t_1)
\label{eq:rectsolnmu} \\
\nu &=& \nu_0\sin\Omega_{20}(t-t_2)
\label{eq:rectsolnnu}
\end{eqnarray}
which is mathematically equivalent to the solution
(\ref{eq:pertSHMx})--(\ref{eq:pertSHMy}) of the simple  harmonic oscillator.
Clearly, $\mu_0\ne0$ and $\nu_0=0$ yields pure rocking motion
in one principal direction, while
 $\mu_0=0$ and $\nu_0\ne0$ yields pure rocking in another.
The case of a central orbit, $\mu_0=\nu_0$ is of particular interest.
The angular momentum quantity
$K=(\mu\dot\nu-\nu\dot\mu)+k(\mu^2+\nu^2)$
(which we shall see to be constant when $\epsilon=0$)
is easily shown to have two components
\[
K = \mu_0^2 \bar\Omega\cos(2\Omega^\prime t-\rho) + 
    \mu_0^2 \Omega^\prime\cos(2\bar\Omega t-\sigma)
\]
where $\bar\Omega=\half(\Omega_{10}+\Omega_{20})$ and
$\Omega^\prime=\half(\Omega_{10}-\Omega_{20})$.
The first component is of large amplitude and low frequency,
the second is of small amplitude and high frequency.
Generically, the trajectory of the solution densely fills
a square region in the $\mu$--$\nu$-plane.



\section{The Routh Sphere $\mathbf{(\IA=\IB)}$:
         complete solution for small amplitude}
\label{sec:RouthSphere}


Let us now consider the solution for 
the Routh Sphere, the symmetric case with $\IA=\IB$ or $\epsilon=0$,
so that $\kappa_{12}=\kappa_{21}=\kappa$,
$k_1=k_2=k$ and $\tilde\Omega_1^2=\tilde\Omega_2^2=\tilde\Omega^2$. 
Then the (positive) eigenvalues for the system (\ref{eq:eeeqn}) are
$$
\beta_{1,2} = \mp k + \sqrt{k^2+\tilde\Omega^2}
$$
where $k= \half(1-\kappa)\omega_3$, and the eigenvectors are 
$(1,1)^{\rm T}$ and $(1,-1)^{\rm T}$.  The general solution is
\begin{eqnarray}
\mu &=& \mu_1 \cos\beta_1(t-t_1) + \mu_2 \cos\beta_2(t-t_2)
\label{eq:RSmu} \\
\nu &=& \mu_1 \sin\beta_1(t-t_1) - \mu_2 \sin\beta_2(t-t_2)
\label{eq:RSnu}
\end{eqnarray}
It follows immediately that
\begin{eqnarray}
\mu^2+\nu^2 &=& 
\mu_1^2 + \mu_2^2 + 2\mu_1\mu_2\cos[(\beta_1+\beta_2)t-b_{12}] 
\label{eq:musqnusq}
\\
\mu\dot\nu-\nu\dot\mu &=& 
\mu_1^2\beta_1 - \mu_2^2\beta_2 +
\mu_1\mu_2(\beta_1-\beta_2)\cos[(\beta_1+\beta_2)t-b_{12}]
\nonumber \\
\mu\dot\mu+\nu\dot\nu &=& 
- \mu_1\mu_2(\beta_1+\beta_2)\sin[(\beta_1+\beta_2)t-b_{12}]
\nonumber
\end{eqnarray}
where $b_{12}=\beta_1t_1+\beta_2t_2$ is a fixed phase.
When the third of these quantities, namely $\mu\dot\mu+\nu\dot\nu$, vanishes,
$\dot\theta=0$ and $\theta$ reaches an
extremum. This occurs when
$$
(\beta_1+\beta_2)t - b_{12} = n\pi
\qquad\mbox{or}\qquad
t = t_n^0 \equiv \frac{ n\pi + b_{12}}{\beta_1+\beta_2}
$$
The cosine factors then take the value $(-1)^n$. We find that
$$
\dot\phi_{2n} = \omega_3 + \frac{\mu_1\beta_1-\mu_2\beta_2}{\mu_1+\mu_2}
\qquad
\dot\phi_{2n+1} = \omega_3 + \frac{\mu_1\beta_1+\mu_2\beta_2}{\mu_1-\mu_2}
$$
Note that both 
$\dot\phi_{2n}$ and $\dot\phi_{2n+1}$ are independent of $n$. If they are of the
same sign, the azimuthal angle $\phi$ changes monotonically with time. If not,
the body executes a reverse loop in each cycle. However, there is
\emph{no recession in either case}; this would require $\dot\phi_n$ to be
a function of $n$.

The absence of recession also follows immediately from
(\ref{eq:musqnusq}). This implies that the distance from the
origin of the $\mu$--$\nu$-plane varies between
$\big||\mu_1|-|\mu_2|\big|$ and $|\mu_1|+|\mu_2|$. So, unless
$|\mu_1|=|\mu_2|$, the accessible region is annular, and
the angular momentum about $\mu=\nu=0$ cannot change sign.

It is straightforward to show that the equations of the Routh Sphere
have two constants, the energy quantity
\[
E_{\mu+\nu} = \half(\dot\mu^2+\dot\nu^2) 
            + \half\tilde\Omega^2(\mu^2+\nu^2) 
\]
and a quantity relating to angular momentum,
\[
K = (\mu\dot\nu-\nu\dot\mu) + k(\mu^2+\nu^2) 
\]
This quantity may be written in terms of the initial conditions
\begin{equation}
K = \sqrt{k^2+\tilde\Omega^2}(\mu_1^2-\mu_2^2)
\label{eq:Kmumu}
\end{equation}
This is interesting, as it allows us to characterise
the solution in terms of the relative sizes of 
$\mu_1$ and $\mu_2$.

We recall from \cite{L&B09} that the Routh Sphere has
two constants in addition to the energy, Jellett's
constant and Routh's constant:
\[
Q_J = \IA s^2 \dot\phi + \IC f \omega_3
\qquad\mbox{and}\qquad
Q_R = \omega_3/\rho
\]
where the density $\rho$ is defined by 
$\rho = 1/\sqrt{\IC + s^2 + (\IC/\IA)f^2}$.
To $O(\theta^2)$, these may be written
\begin{eqnarray*}
\tilde Q_J &=& (\IA\theta^2\dot\phi+\IC f_0\omega_3)
             - \half \IC\omega_3\theta^2
\\
\tilde Q_R &=& \left[ 1 + 
        \left(\frac{\IA-\IC f_0}{(\IA+f_0^2)\IC}\right)
        \frac{\theta^2}{2}\right] \frac{\omega_3}{\rho_0}
\end{eqnarray*}
where $\rho_0=1/\sqrt{(\IA+f_0^2)\IC/\IA}$.  We can now show that
\begin{equation}
K = \frac{1}{4\IA}\left[\tilde Q_J-\IC f_0\rho_0\tilde Q_R\right]
\label{eq:Kconst}
\end{equation}
This allows us to relate the solution in terms of $\mu_1$ and $\mu_2$
(or $K$), to the quantities $\tilde Q_J$ and $\tilde Q_R$
(\emph{cf.} Fig.~4 in \cite{L&B09}).


\subsection{Epicycle character of the solution}

We will first interpret the solution in the $\mu$--$\nu$-plane,
noting that it bears a correspondence
to the $\theta$--$\phi$-plane through the relationships
\[
\theta = 2\sqrt{\mu^2+\nu^2} \,, \qquad
\phi = \omega_3 t + \arctan{(\nu/\mu)}
\]
We note that the solution (\ref{eq:RSmu})--(\ref{eq:RSnu}) is
comprised of two components: the first
\[
\mu = \mu_1 \cos[\beta_1(t-t_1)] \,,
\qquad
\nu = \mu_1 \sin[\beta_1(t-t_1)] 
\]
is a counterclockwise circular motion with frequency
$\beta_1$ and radius $\mu_1$; the second
\[
\mu =  \mu_2 \cos[\beta_2(t-t_2)] \,,
\qquad
\nu = -\mu_2 \sin[\beta_2(t-t_2)] 
\]
is a clockwise circular motion with frequency
$\beta_2$ and radius $\mu_2$. The complete motion
is thus \emph{an epicycle}.

\begin{figure}
\begin{center}
\includegraphics[scale=0.275]{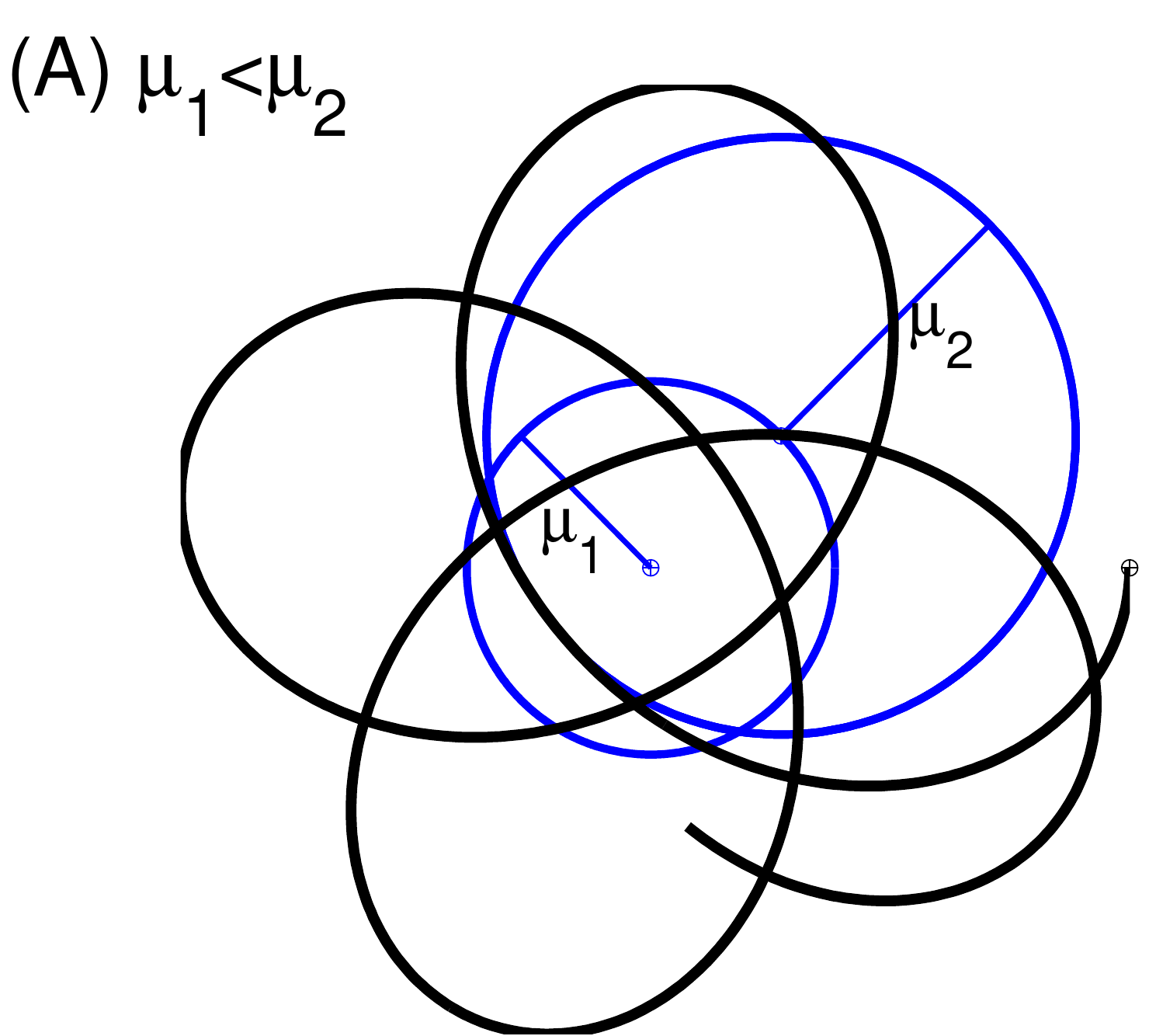}
\hspace{5mm}
\includegraphics[scale=0.275]{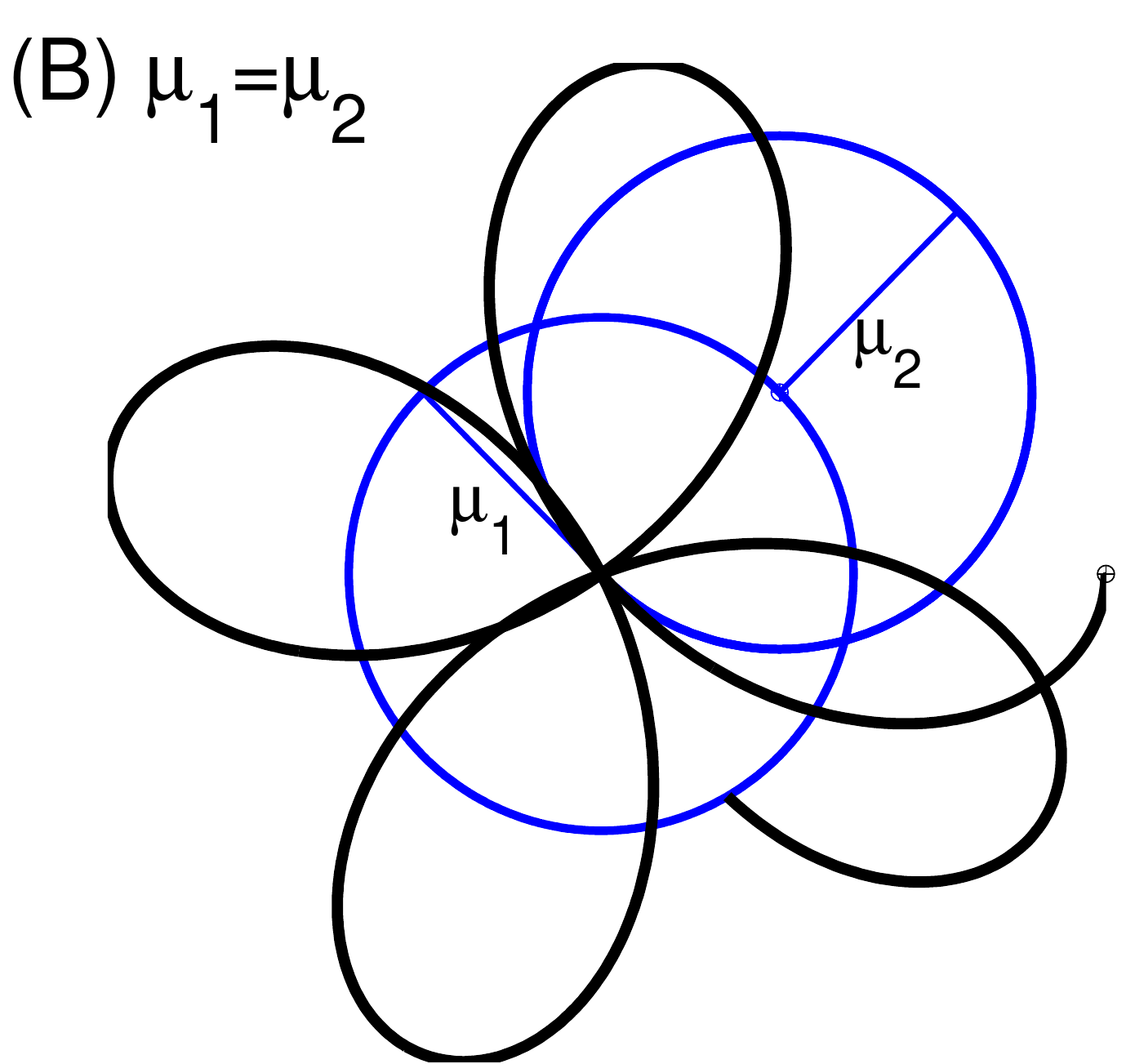}
\hspace{5mm}
\includegraphics[scale=0.275]{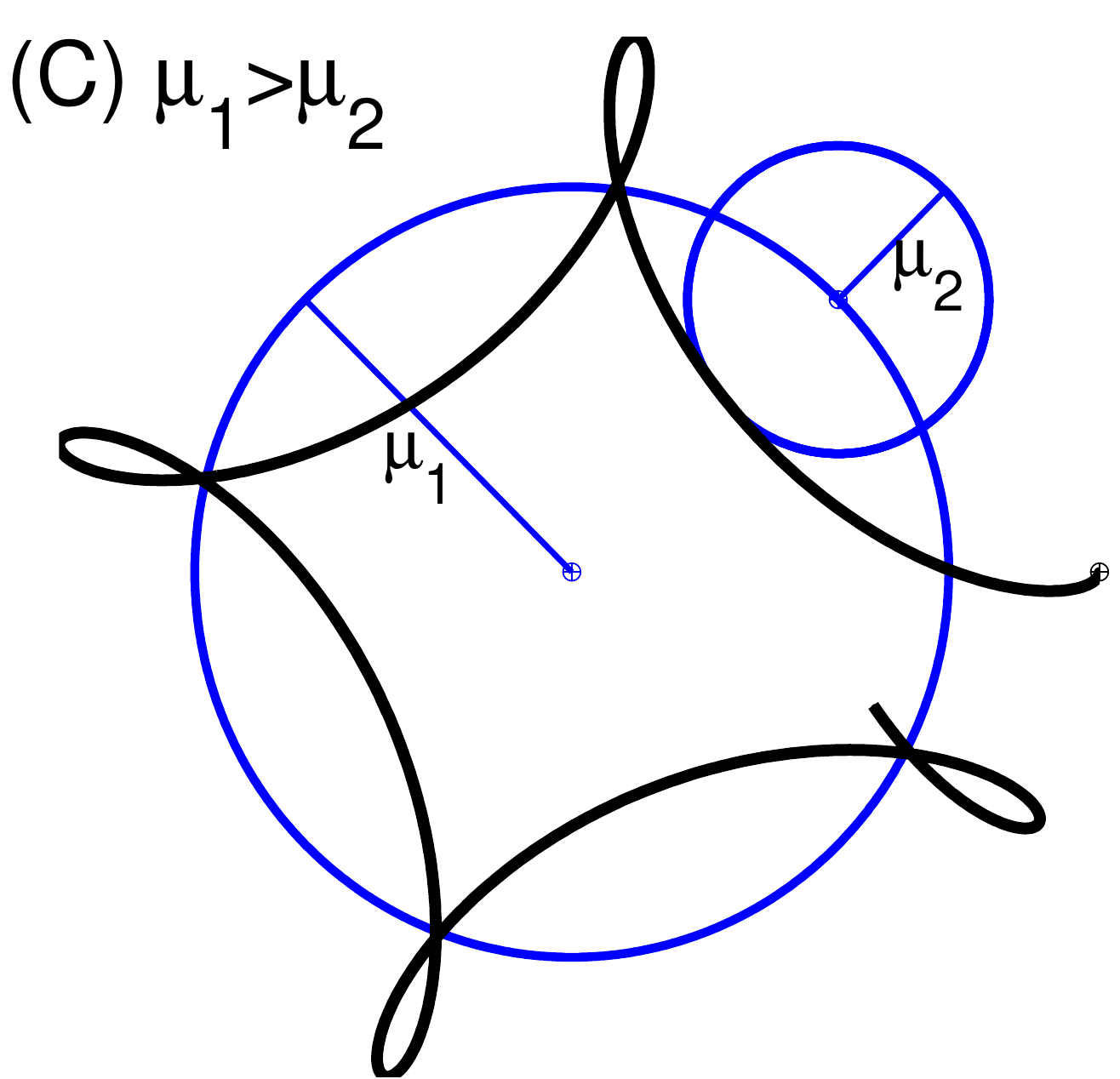}
\caption{Routh Sphere: the trajectories in the
$\mu$--$\nu$-plane are epicycles.
The three distinct cases are illustrated:
left panel $\mu_1<\mu_2$ (centripetal orbit in $\mu$--$\nu$-plane);
centre panel $\mu_1=\mu_2$ (central orbit);
right panel $\mu_1>\mu_2$ (centrifugal orbit in $\mu$--$\nu$-plane).
The frequencies are
$\beta_1=0.3$ (counterclockwise) and $\beta_2=1.0$ (clockwise).}
\label{fig:epicycles}
\end{center}
\end{figure}

We illustrate various possibilities schematically
in Fig.~\ref{fig:epicycles}.
For $\mu_1 < \mu_2$, the orbit circulates about the centre, $\mu=\nu=0$,
always curving towards it in a centripetal orbit (Fig.~\ref{fig:epicycles}(A)).
For $\mu_1 > \mu_2$, the orbit circulates in the opposite direction
about the centre, sometimes curving towards it and sometimes curving away.
This is a centrifugal orbit (Fig.~\ref{fig:epicycles}(C)).
For $\mu_1 = \mu_2$, the orbit passes periodically through
the centre $\mu=\nu=0$.
We call this a central orbit (Fig.~\ref{fig:epicycles}(B));
for further discussion, see Appendix~B.

We have chosen $0\le\beta_1\le\beta_2$ by arbitrary convention.
The solution for $\mu$ and $\nu$ is given by
(\ref{eq:RSmu})--(\ref{eq:RSnu}). To transform back to
the $\theta$--$\phi$-plane, we must rotate through an angle
$\omega_3 t$.
Defining $x=(\theta/2)\cos\phi, y=(\theta/2)\sin\phi$, we have
\begin{eqnarray*}
x &=& \mu_1\cos(\alpha_1 t - \beta_1 t_1 )
    + \mu_2\cos(\alpha_2 t - \beta_2 t_1 ) \\
y &=& \mu_1\sin(\alpha_1 t - \beta_1 t_1 )
    - \mu_2\sin(\alpha_2 t - \beta_2 t_1 ) 
\end{eqnarray*}
where the frequencies are
$\alpha_1 = \beta_1 + \omega_3$ and $\alpha_2 = \beta_2 - \omega_3$.
We find that $0\le\alpha_2\le\alpha_1$. This effectively switches
the roles of the two components of the solution: 
centripetal motion in the $\mu$--$\nu$-plane corresponds to centrifugal
in the $\theta$--$\phi$-plane, and vice versa.  Referring to (\ref{eq:Kmumu})
and (\ref{eq:Kconst}), we see that the following correspondence holds:
\begin{eqnarray*}
{\ }\hspace{-15mm}
[\mu_1 > \mu_2]
\Longleftrightarrow
[K > 0]
&\Longleftrightarrow
&[\tilde Q_J > \tilde Q_{J,0}^{\rm crit}]
\Longleftrightarrow
[\mbox{\rm Centripetal (in}\ \theta\mbox{--}\phi\mbox{-plane})]
\cr                     
{\ }\hspace{-15mm}
[\mu_1 = \mu_2]
\Longleftrightarrow
[K = 0]
&\Longleftrightarrow
&[\tilde Q_J = \tilde Q_{J,0}^{\rm crit}]
\Longleftrightarrow
[\mbox{\rm Central (in}\ \theta\mbox{--}\phi\mbox{-plane})]
\cr                     
{\ }\hspace{-15mm}
[\mu_1 < \mu_2]
\Longleftrightarrow
[K < 0]
&\Longleftrightarrow
&[\tilde Q_J < \tilde Q_{J,0}^{\rm crit}] 
\Longleftrightarrow
[\mbox{\rm Centrifugal (in}\ \theta\mbox{--}\phi\mbox{-plane})]
\end{eqnarray*}
where $\tilde Q_{J,0}^{\rm crit} = \IC f_0\rho_0\tilde Q_R$
(see also Fig.~4 in \cite{L&B09}).

\begin{figure}
\includegraphics[scale=0.45]{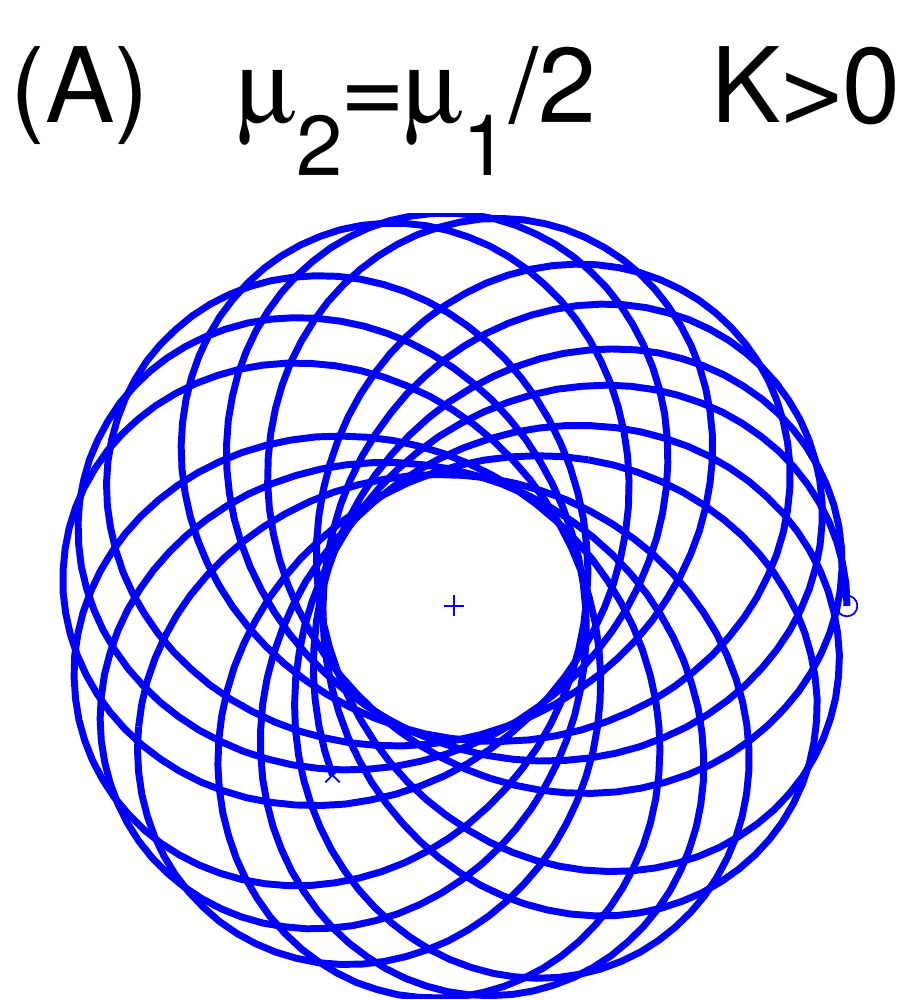}
\hspace{5mm}
\includegraphics[scale=0.45]{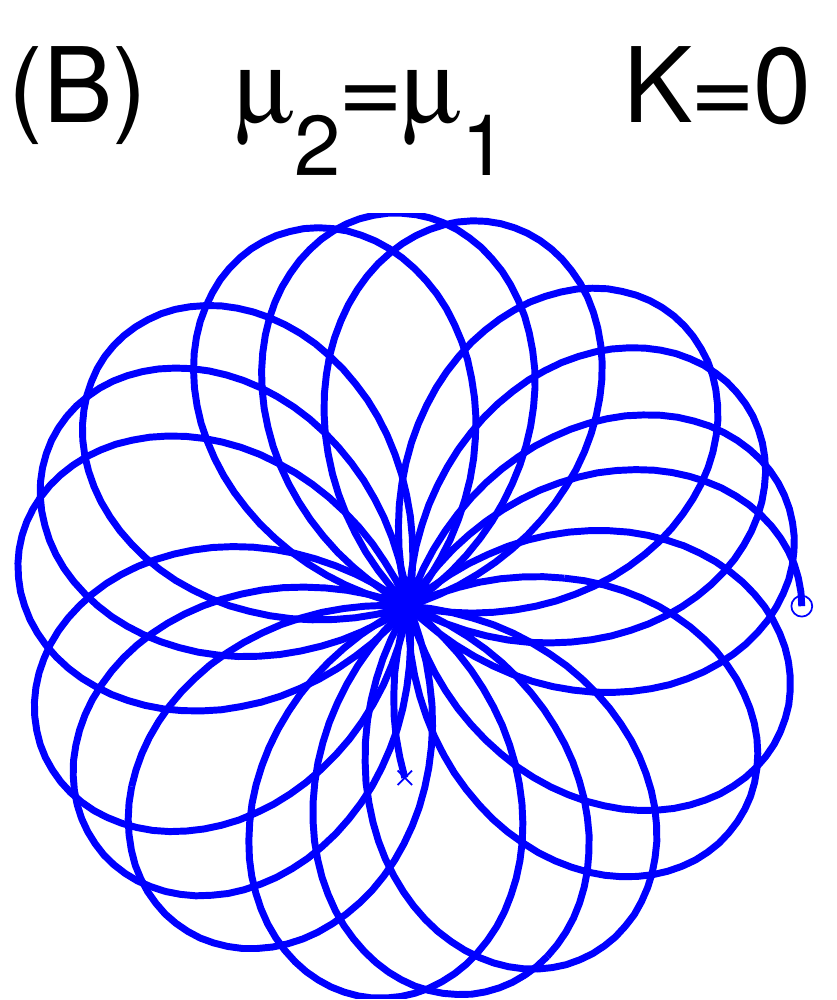}
\hspace{5mm}
\includegraphics[scale=0.45]{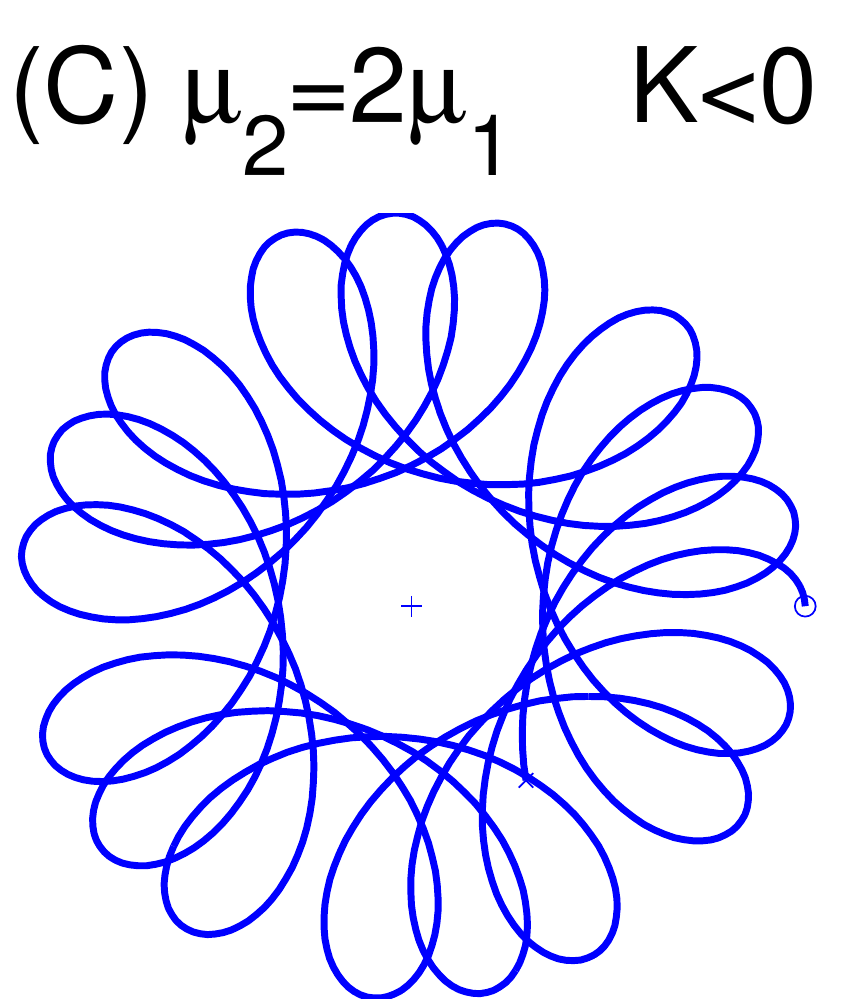}

\includegraphics[scale=0.45]{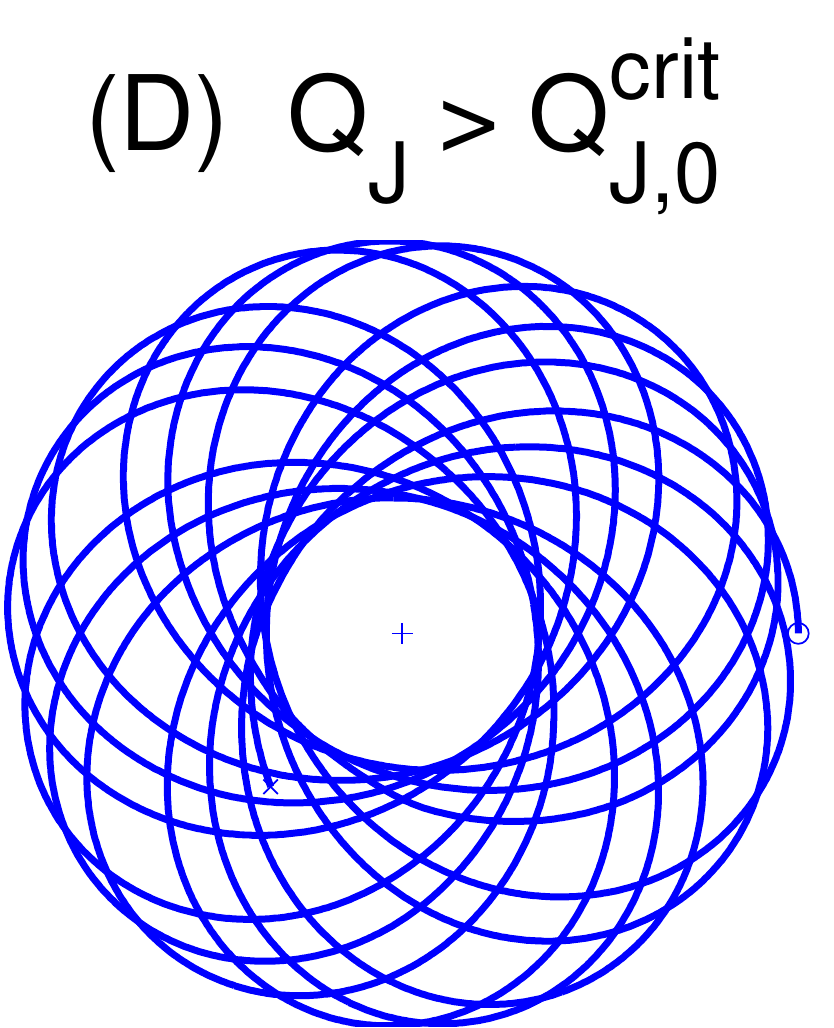}
\hspace{7mm}
\includegraphics[scale=0.45]{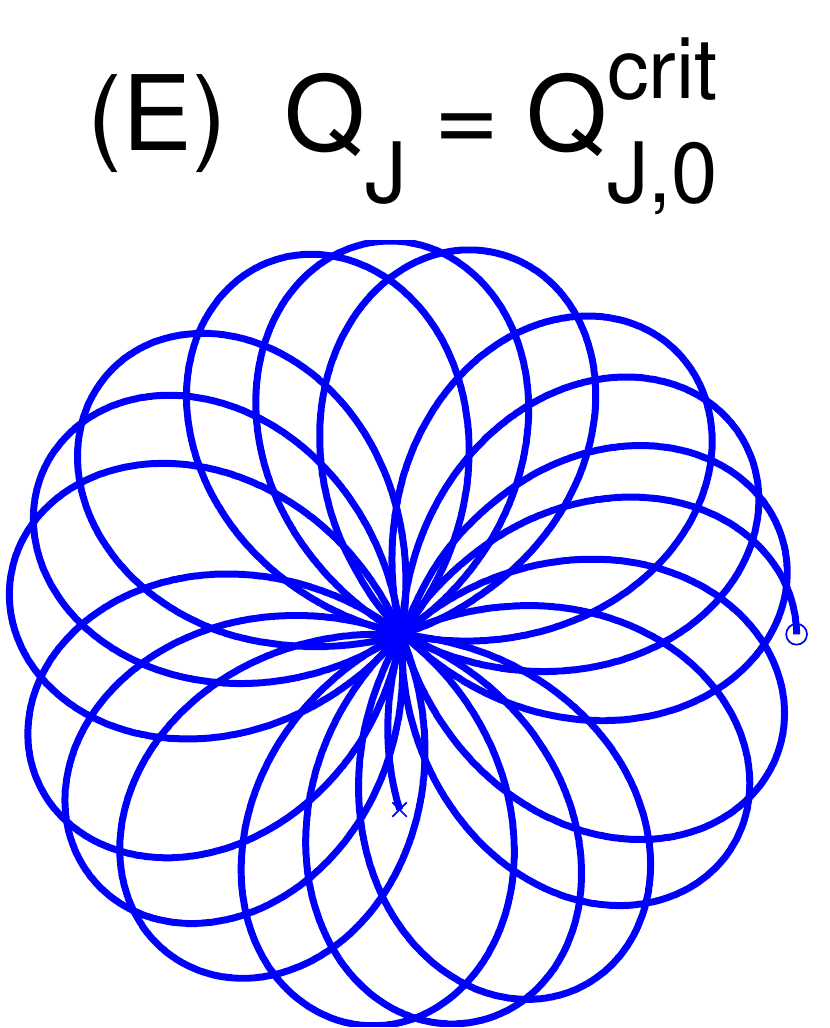}
\hspace{7mm}
\includegraphics[scale=0.45]{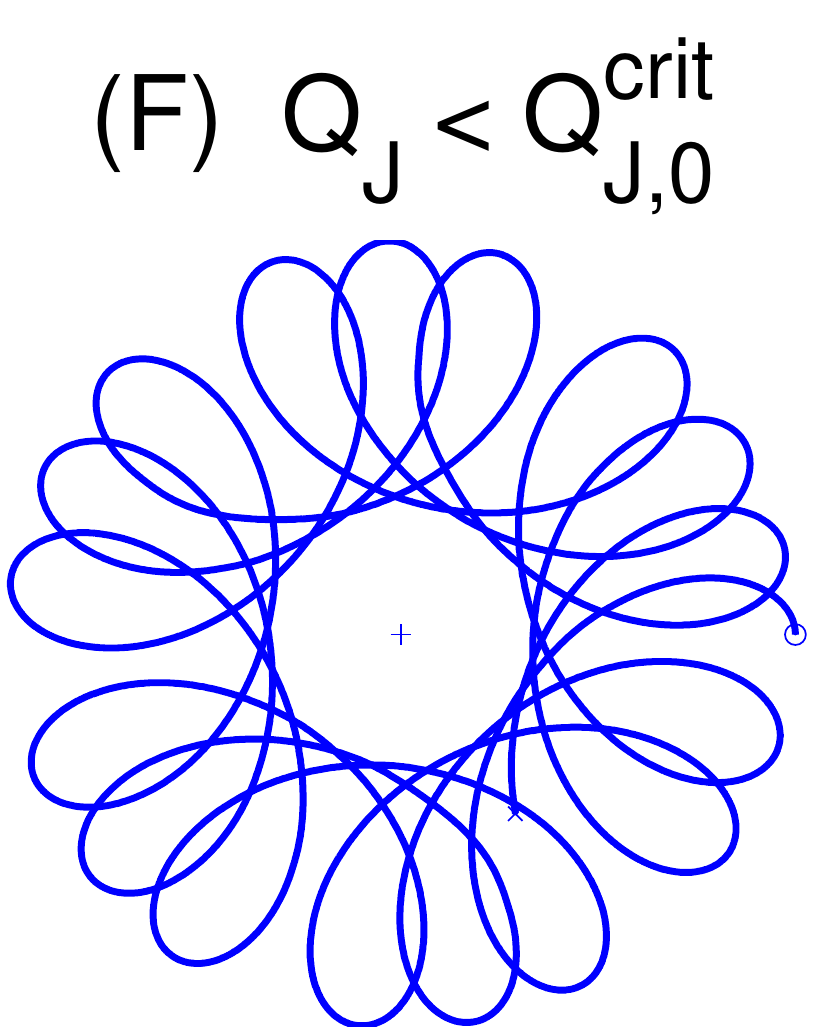}

\caption{Routh Sphere:
the trajectories in the $\theta$--$\phi$-plane are epicycles.
Panels (A)--(C): Analytical solutions of the quaternion equations
for three sets of initial conditions:
(A) $\mu_2=0.5\mu_1$; (B) $\mu_2=\mu_1$; (C) $\mu_2=2\mu_1$.
In all cases, $\mu_1=1$ and $t_1=t_2=0$.
Panels (D)--(F): Numerical solutions of the full nonlinear equations for
the corresponding initial conditions:
(D) $\omega_2(0)=0.008147$; (E) $\omega_2(0)=0.005389$; (F) $\omega_2(0)=0.002632$.
In all cases $\theta(0)=0.01$, $\omega_3(0)=1.0$ and $\phi(0)=\psi(0)=\omega_1(0)=0$.}

\label{fig:Qpanels}
\end{figure}

\begin{figure}
\includegraphics[height=45mm]{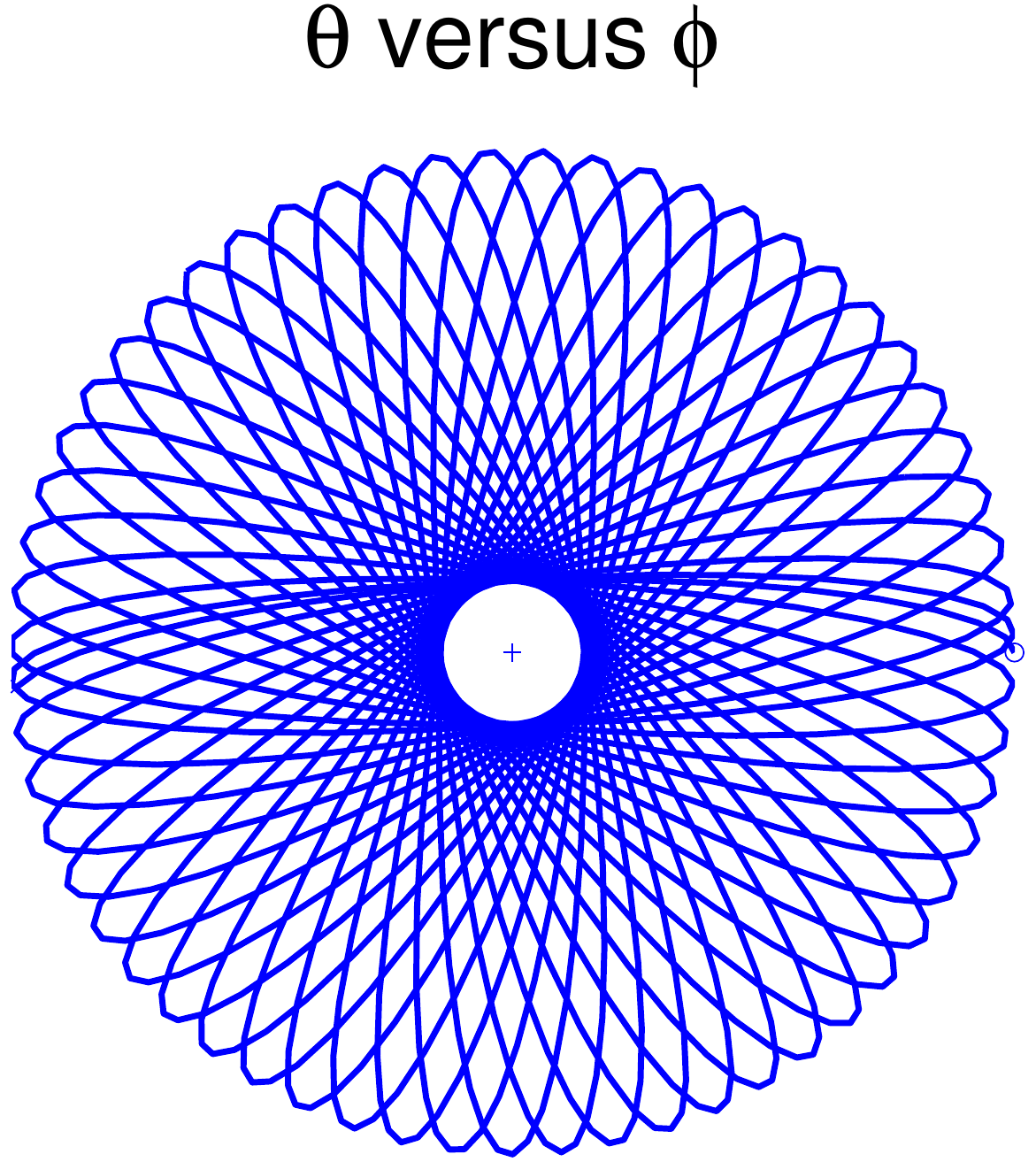}
\hspace{5mm}
\includegraphics[height=45mm]{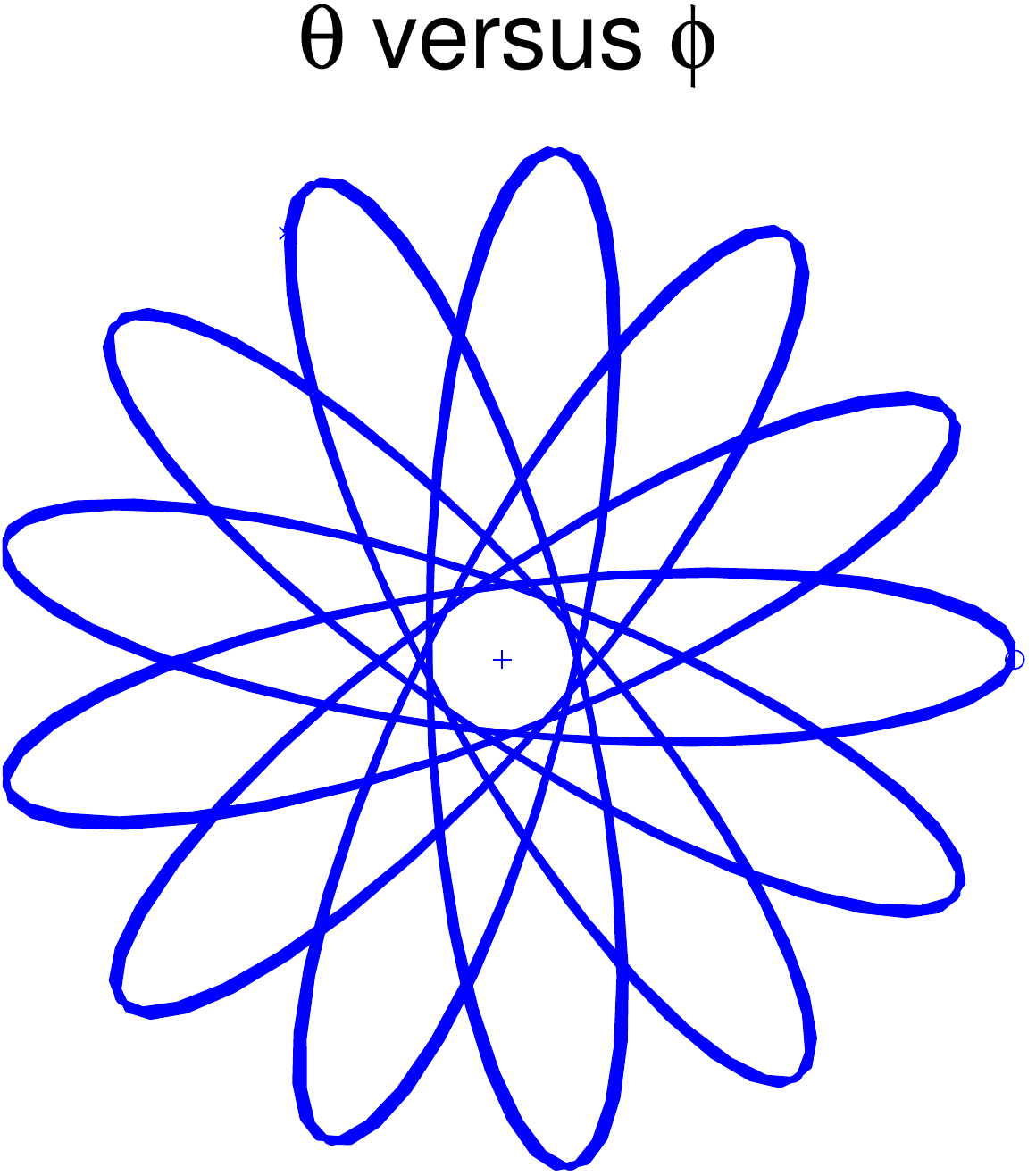}
\hspace{5mm}
\includegraphics[height=45mm]{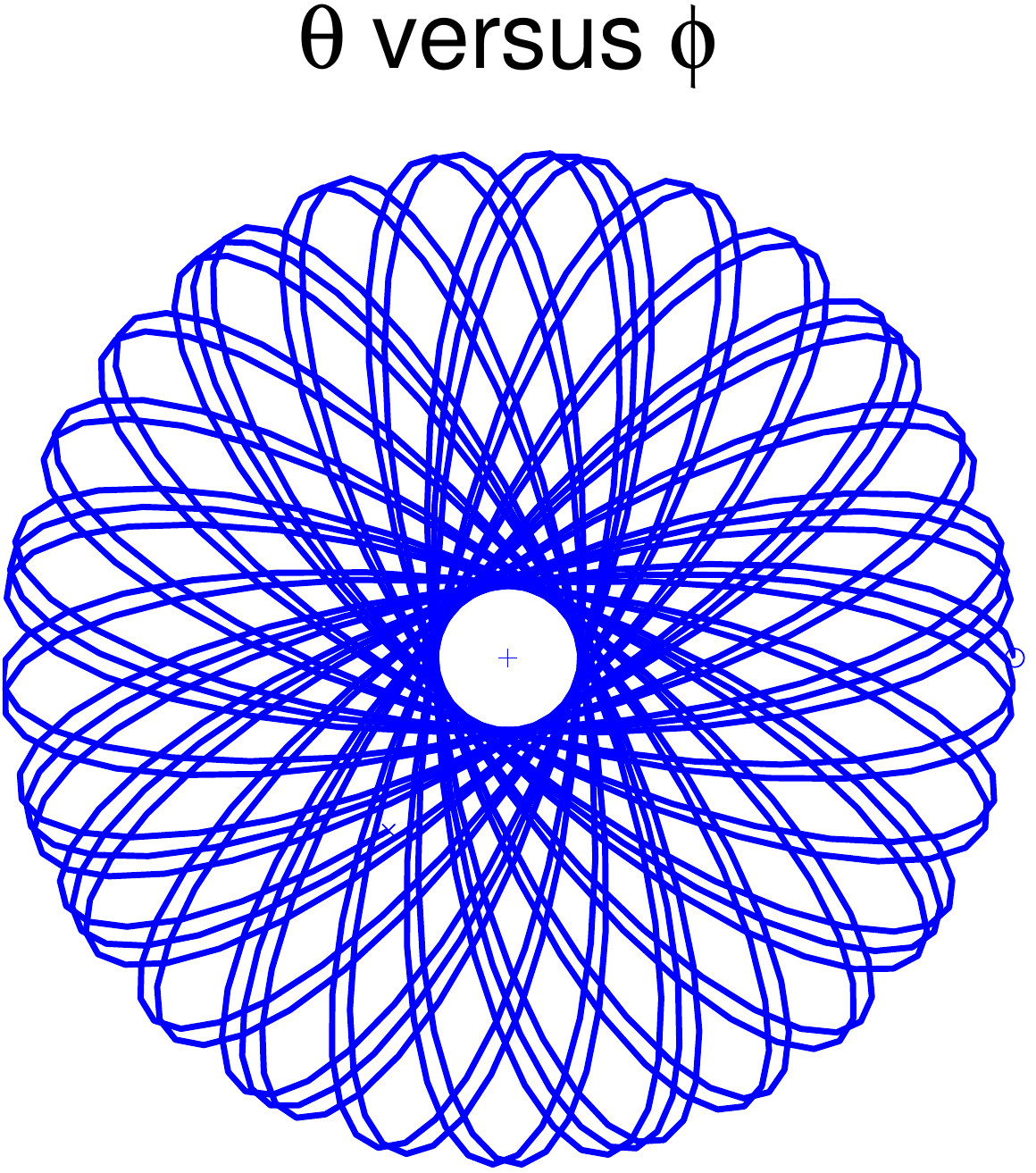}

\includegraphics[height=45mm]{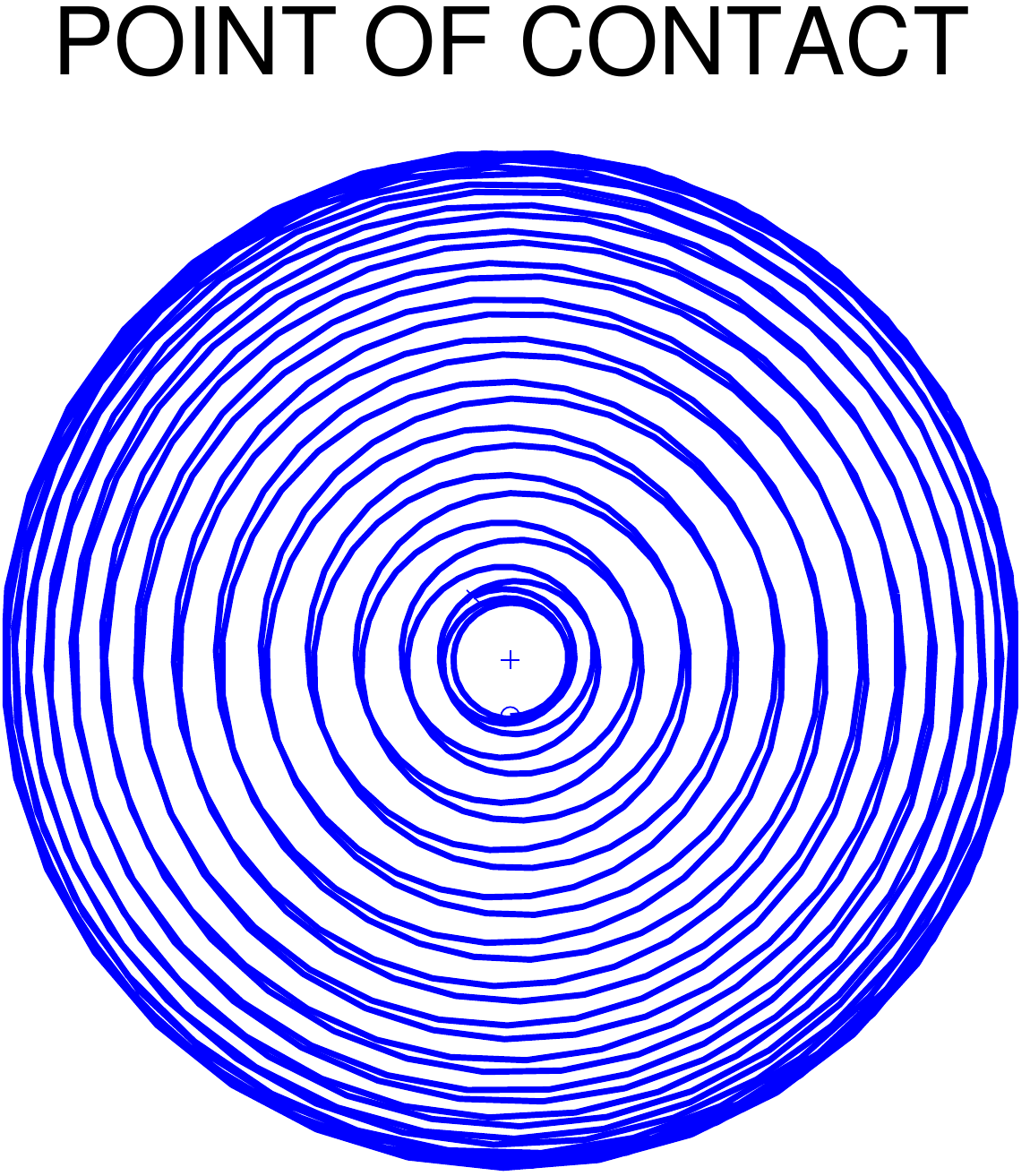}
\hspace{5mm}
\includegraphics[height=45mm]{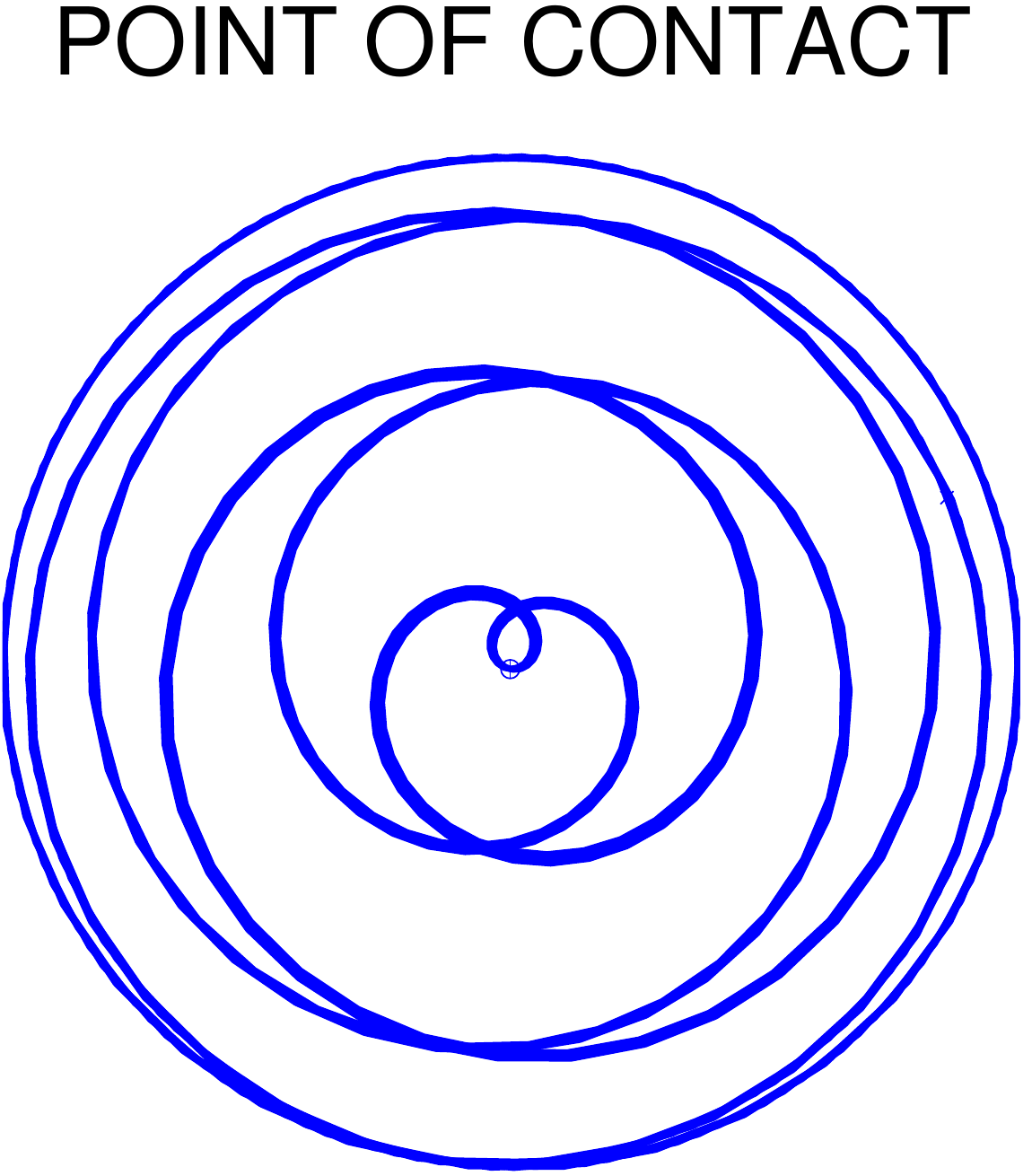}
\hspace{5mm}
\includegraphics[height=45mm]{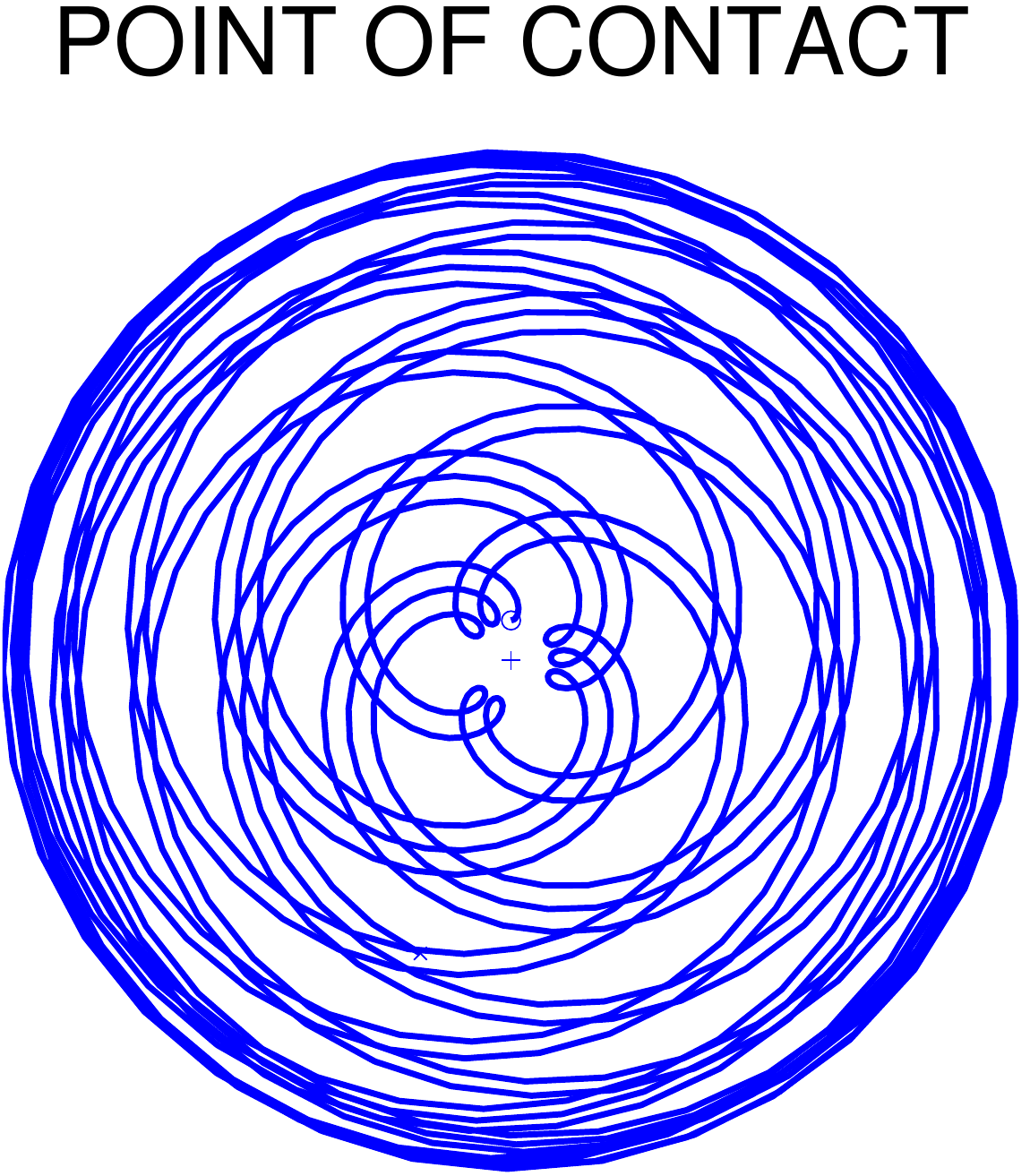}

\caption{Top row: trajectories in the $\theta$--$\phi$-plane for three sets
of initial conditions,
$\omega_2=0.001005$, $\omega_3(0)=0.02$ (left),
$\omega_2=0.001384$, $\omega_3(0)=0.09$ (centre) and
$\omega_2=0.001658$, $\omega_3(0)=0.14$ (right).
Bottom row: corresponding plots of the trajectory of the point of contact.
In all cases $\theta(0)=0.01$, $\phi(0)=0$, $\psi(0)=0$ and $\omega_1(0)=0$
and $\mu_2/\mu_1=\frac{3}{4}$. The solutions in the central column satisfy the
criterion (\ref{eq:centralPOC}).}

\label{fig:POC}
\end{figure}


\subsection{Trajectory of the point of contact}

So far, we have looked at the projections of the orbit
in the $\mu$--$\nu$-plane and in the $\theta$--$\phi$-plane.
However, the external observer is aware of both the
orientation of the body, as determined by the Euler angles and the position
as given by the coordinates $(X,Y)$ of the geometric centre or, equivalently,
the point of contact.

The movement of the geometric centre is linked to the angular velocity of the
body through the rolling constraint \cite{L&B09}:
\[
(\dot X,\dot Y,0) = \mathbf{\boldsymbol\omega\boldsymbol\times K}
\]
where $\mathbf{K}$ is a unit vertical vector.
In terms of quaternions, this becomes
\begin{eqnarray*}
\dot X &=& 2[\gamma\dot\eta-\eta\dot\gamma+\zeta\dot\xi-\xi\dot\zeta] \\
\dot Y &=& 2[\xi\dot\gamma-\gamma\dot\xi+\zeta\dot\eta-\eta\dot\zeta] 
\end{eqnarray*}
Substituting the solutions 
(\ref{eq:gammazeta}), (\ref{eq:gensocone}) and (\ref{eq:gensoctwo})
for the quaternion components, we get
\begin{eqnarray}
X &=& \phantom{-.}
            \left[{2\mu_1\beta_1}/{\alpha_1}\right]\sin(\alpha_1 t-\beta_1 t_1) 
           -\left[{2\mu_2\beta_2}/{\alpha_2}\right]\sin(\alpha_2 t-\beta_2 t_2) 
\label{eq:XPOC} \\
Y &=& -\left[{2\mu_1\beta_1}/{\alpha_1}\right]\cos(\alpha_1 t-\beta_1 t_1) 
           -\left[{2\mu_2\beta_2}/{\alpha_2}\right]\cos(\alpha_2 t-\beta_2 t_2)
\label{eq:YPOC}
\end{eqnarray}
We saw that the criterion for a central orbit,
or the boundary between a centripetal and a centrifugal orbit,
in the $\theta$--$\phi$-plane was $\mu_1=\mu_2$.
The corresponding boundary for the $X$--$Y$-plane is the equality of the 
coefficients in (\ref{eq:XPOC})--(\ref{eq:YPOC}) or
\begin{equation}
\frac{\mu_1\beta_1}{\alpha_1} = \frac{\mu_2\beta_2}{\alpha_2} 
\label{eq:centralPOC}
\end{equation}
The distinction is a reflection of the nonholonomic nature of the constraint:
we cannot express $(X,Y)$ in terms of the Euler angles until the solution
is found.

In Fig.~\ref{fig:POC} we show trajectories in the $\theta$--$\phi$-plane
(top row) and corresponding plots for the point of contact (bottom row).
In the three cases, the orbit is centripetal in the $\theta$--$\phi$-plane
but in the $X$--$Y$-plane it changes character, from centripetal to central
to centrifugal as $\omega_3$ increases.  The solutions in the central
column of Fig.~\ref{fig:POC} satisfy the criterion (\ref{eq:centralPOC}).





\section{Conclusion}
\label{sec:Conclusion}

Box and loop orbits are found in a wide range
of physical systems. We illustrate them in the elementary
context of a perturbed simple harmonic oscillator.
Then, the dynamical equations for small
amplitude motions of the \RnR\ are expressed in terms of
quaternions. The complete solution is expressed as an
\emph{epi-ellipse}, a combination of two purely elliptic motions.
This allows us to clarify the phenomenon of recession, and the
conditions under which it occurs.
In the particular case of a symmetric body ($\epsilon=0$), the
Routh Sphere, the solution reduces to an epicycle. Only loop
orbits occur and there is no recession.

We have confined attention in the present study to the
dynamics at first order in the polar angle $\theta$.
In an extension of this work, we will present a more detailed
perturbation analysis, including a rigorous demonstration of
energy conservation to second order, explicit expressions
for the Routh and Jellett quantities $Q_R$ and $Q_J$
and a complete analysis of the recession of the \RnR.

The dynamics of the rattleback or celt have been discussed in
many publications; see, for example, \cite{BoKiMa06}. 
It is an ellipsoidal body that exhibits a variety of
reversals of rotation.  While the dominant behaviour of the
rattleback is due to the mis-allignment of its inertial and
geometric axes, the mechanism of recession described here for
the \RnR\ must also be present, and may be proposed as a 
mechanism accounting for observed multiple reversals of the
rattleback. This speculation deserves further consideration.

One of the motivations for studying the \RnR\ is the hope of
finding an invariant of the motion in addition to the energy.
This expectation arises from the symmetry of the body.
For the general loaded sphere, there is a finite
angle $\bdelta$ between the principal
axis corresponding to $\IC$ and the line joining the centres
of gravity and symmetry. For the \RnR, this angle is zero
and the Lagrangian is independent of the azimuthal angle
$\phi$. However, we have not found a second invariant and,
considering the non-holonomic nature of the problem, its
existence remains an open question.


\section*{Appendix A: Euler Angle Ambiguity}
\label{sec:appendix-A}



In his work on celestial mechanics, Euler showed that any
two independent orthogonal frames can be related by
(not more than) three rotations about the coordinate axes.
Kuipers \cite{Kuipers99}
lists twelve sequences of rotations.
The first, denoted $xyz$, means a rotation about the
$x$-axis, followed by a rotation about the new $y$-axis
followed by a rotation about the newer $z$-axis. 
Different choices are made in different areas of science,
frequently leading to ambiguity in the meaning of 
the Euler angles.
In mechanics, two sequences are in common use,
each commanding the respect due to its adoption by
renowned authorities.

We denote the unit orthogonal triad  in the space frame by
$(\mathbf{I,J,K})$ and the corresponding triad in the body
frame by $(\mathbf{i,j,k})$. Coordinates in the space frame
are $(X,Y,Z)$ and in the body frame $(x,y,z)$. The origin is
colocated in the two frames. The  body frame may be related to 
the space frame by a set of three rotations.
In both rotation sequences, the first rotation is about the (space)
$Z$ axis (about the vector $\mathbf{K}$),
and the third is about the (body) $z$ axis
(about the vector $\mathbf{k}$).
However, the axis of the second rotation differs in the 
two sequences and, as a result, the magnitudes of the
rotations also differ.

The first sequence is the $zxz$-sequence, and we denote the
Euler angles in this case by $(\phi,\theta,\psi)$.
In this $zxz$-sequence, favoured by
Landau \&\ Lifshitz \cite{LandauLifshitz76},
Arnold \cite{Arnold78}
and Goldstein {\em et al.} \cite{GoldsteinPooleSafko02},
the second rotation is of an angle $\theta$ about the $x$-axis
that results from the first rotation.
In the second sequence, $zyz$, employed by
Whittaker \cite{Whittaker37}
and by Synge and Griffith \cite{SyngeGriffith59},
we denote the angles by $(\Phi,\Theta,\Psi)$. 
The second rotation is now through an angle $\Theta$ about the
$y$-axis that results from the first rotation.
The overall rotation must be identical for the two sequences.
Constructing the rotation matrix for the composition of the
three rotations in each case and equating the two results, we
find that the relationship between the Euler anges in the
two sequences is
$$
\Phi = \phi-\frac{\pi}{2} \qquad
\Theta = \theta           \qquad
\Psi = \psi+\frac{\pi}{2} \,.
$$
These relationships enable us to convert between the two conventions.
The full rotation matrix for $zxz$ is given, for example, in
Goldstein {\em et al.} \cite[pg.~153]{GoldsteinPooleSafko02}, 
and in Marsden and Ratiu \cite[pg.~494]{MarsdenRatiu99}.
For the $zyz$ sequence, the rotation matrix is given in
Whittaker \cite[pg.~10]{Whittaker37}
and in Synge and Griffith \cite[pg.~261]{SyngeGriffith59}.

The quaternion representing the rotation must be independent
of the Euler angle convention. However, the expressions for the
quaternion components in terms of the angles will be different in
each case. This explains why our definitions of the components
$(\gamma,\xi,\eta,\zeta)$ are different from those of
$(\chi,\xi,\eta,\zeta)$ in Whittaker
\cite{Whittaker37}.
The expansion of the components of angular velocity
$(\omega_1,\omega_2,\omega_3)$ in terms of angles is also
different in the two conventions, but their expression in
terms of  $(\gamma,\xi,\eta,\zeta)$ is identical. Thus, many
of the formulae we derive are the same as those found
in \cite{Whittaker37}, except that we replace $\chi$ by $\gamma$
to avoid confusion with our notation for $\cos\psi$.

Altmann \cite{Altmann86}
observes that sequence $zyz$ is now universal in quantum physics,
as it is consistent with the Condon and Shortley convention.
In the context of classical mechanics, there is no obvious advantage
of either convention over the other.
However, we feel that it is important to avoid any ambiguity by making
the choice clear. Sequence $zxz$ is used in the present paper.


\section*{Appendix B: Epicyclic and epi-elliptic orbits}
\label{sec:appendix-B}



\subsection*{Epicyclic motion}

\begin{figure}
\begin{center}
\includegraphics[scale=0.35]{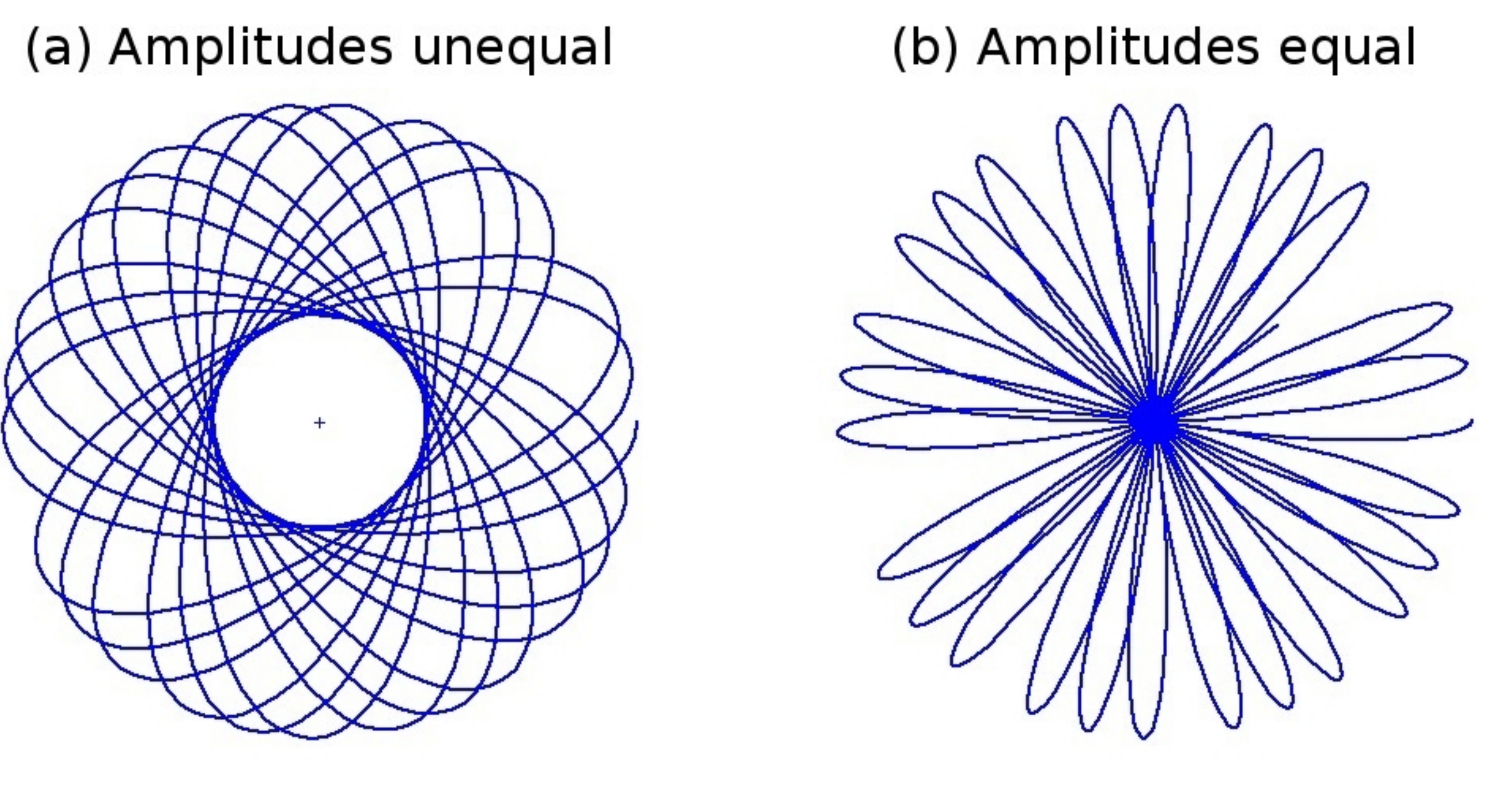}
\caption{Two epicyclic orbits.
(a) Amplitudes unequal: $\mu_1=1.0$, $\mu_2=2.0$.
(b) Amplitudes equal: $\mu_1=\mu_2=1.0$. 
In both cases, $\beta_1=1.0$ and $\beta_2=\pi^2/8$.}
\label{fig:twoepicycles}
\end{center}
\end{figure}

We consider the character of solutions of the form
\begin{eqnarray}
\mu &= \phantom{\lambda_1}\mu_1\cos\beta_1(t-t_1)
    &+ \phantom{\lambda_2}\mu_2\cos\beta_2(t-t_2) 
\label{eq:gensocmu} \\
\nu &=          \lambda_1 \mu_1\sin\beta_1(t-t_1) 
    &+          \lambda_2 \mu_2\sin\beta_2(t-t_2)
\label{eq:gensocnu}
\end{eqnarray}
as obtained above, (\ref{eq:gensocone})--(\ref{eq:gensoctwo}),
for small-amplitude motions of the \RnR.
For the Routh Sphere ($\epsilon=0$), $\lambda_1=1$ and $\lambda_2=-1$.
The motion consists of two components representing circular
motion in opposite directions. We have chosen the order of
the eigenvalues such that $0\le\beta_1\le\beta_2$. Generically,
$\beta_1$ and $\beta_2$ are incommensurate and the orbit is
dense in an annular region
$\varrho_{\rm min}\le\varrho\le\varrho_{\rm max}$ where 
$\varrho=\sqrt{\mu^2+\nu^2}$, 
$\varrho_{\rm min}=\big||\mu_1|-|\mu_2|\big|$ and
$\varrho_{\rm max}=|\mu_1|+|\mu_2|$.
The trajectory may be centripetal (always curving towards the
origin $\varrho=0$) or centrifugal (sometimes curving away),
but it is always a loop orbit. Precession is particularly
evident when $|\mu_1|\approx|\mu_2|$; 
see Fig.~\ref{fig:twoepicycles}(a).
When $\mu_1=\mu_2$, the solution may be written as a central orbit
\begin{equation}
{\mu  \choose \nu}
= 2\mu_1\cos\beta_{+}(t-t_{+})\cdot
{\cos \choose \sin}
\beta_{-}(t-t_{-})
\nonumber
\end{equation}
where $\beta_{+}=\half(\beta_1+\beta_2)$,
$\beta_{-}=\half(\beta_1-\beta_2)$,
$t_{+}=(\beta_1t_1+\beta_2t_2)/(\beta_1+\beta_2)$ and
$t_{-}=(\beta_1t_1-\beta_2t_2)/(\beta_1-\beta_2)$.
This is a central orbit, which passes twice through the origin
on each cycle (Fig.~\ref{fig:twoepicycles}(b)).
The precession angle is given by
$\Delta\varphi=(\beta_{-}/\beta_{+})2\pi$.

\begin{figure}
\begin{center}
\includegraphics[scale=0.35]{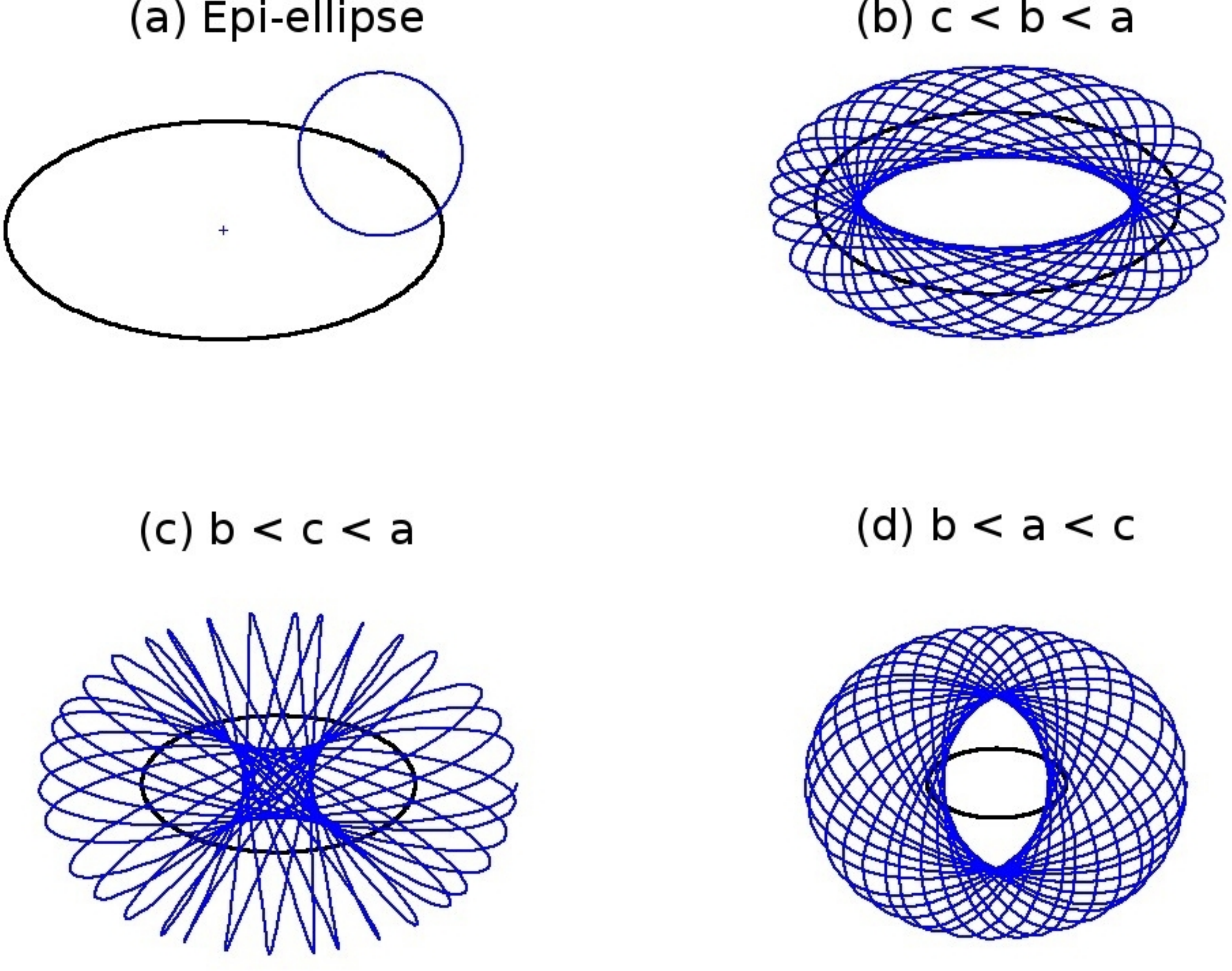}
\caption{
(a) Inner ellipse with major and minor semi-axes $a$ and $b$.
Outer ellipse taken as a circle of radius $c$.
(b) $c<b<a$: loop orbit.
(c) $b<c<a$: box orbit.
(d) $b<a<c$: loop orbit.
In all cases, $\beta_1=1.0$ and $\beta_2=\pi^2/8$.
The inner ellipse is shown as a heavy curve in each panel.}
\label{fig:epiellipses}
\end{center}
\end{figure}

\subsection*{Epi-elliptic motion}

The solution (\ref{eq:gensocmu})--(\ref{eq:gensocnu}) has
two components, each being a trajectory that is elliptic
The first component is
\[
\mu = \mu_1 \cos[\beta_1(t-t_1)] \,,
\qquad
\nu = \nu_1 \sin[\beta_1(t-t_1)]  \,,
\]
elliptical motion with frequency
$\beta_1$ and semi-axes $\mu_1$ and $\nu_1=\mu_1\lambda_1$.
The second component is
\[
\mu =  \mu_2 \cos[\beta_2(t-t_2)] \,,
\qquad
\nu =  \nu_2 \sin[\beta_2(t-t_2)]  \,,
\]
elliptical motion with frequency
$\beta_2$ and semi-axes $\mu_2$ and $\nu_2=\mu_2\lambda_2$.
The character of the trajectory is thus determined by the
relative magnitudes and signs of the parameters
$\{\mu_1,\nu_1,\mu_2,\nu_2\}$.

First, we consider the case where the second ellipse fits
within the first:
\[
|\mu_1|-|\mu_2| > 0
\qquad\mbox{and}\qquad
|\nu_1|-|\nu_2| > 0 \,.
\]
Clearly, $\varrho$ remains positive, with
$\varrho_{\rm min}=\min\{(|\mu_1|-|\mu_2|),(|\nu_1|-|\nu_2|)\}$.
There is an exclusion zone around the origin, inaccessible to
the trajectory and we have a loop orbit.
Next, we consider the case where the first ellipse fits within
the second:
\[
|\mu_1|-|\mu_2| < 0
\qquad\mbox{and}\qquad
|\nu_1|-|\nu_2| < 0 \,.
\]
Again, $\varrho$ is bounded away from zero and
the trajectory is a loop orbit.
The remaining case is where 
$(|\mu_1|-|\mu_2|)$ and $(|\nu_1|-|\nu_2|)$ are of opposite
signs. Then $\varrho$ may become zero, the trajectory may
pass through the origin and the orbit is of box type. Since it
is only in this case that recession may be observed, the
criterion for recession may be written
\begin{equation}
(|\mu_1|-|\mu_2|)\cdot(|\nu_1|-|\nu_2|) < 0 \,.
\label{eq:BLCriteron}
\end{equation}
as stated in (\ref{eq:rescrit}) in \S\ref{sec:RnRasym}.

To illustrate the character of the orbits, we plot some
solutions in Fig~\ref{fig:epiellipses}. We let the major
and minor semi-axes of the inner ellipse be $a$ and $b$
and those of the outer ellipse be $c$ and $d$.
Without loss of generality, we take $c=d$.
Then the three cases discussed above are presented in
panels (b), (c) and (d) of Fig~\ref{fig:epiellipses}.
The criterion (\ref{eq:BLCriteron}) for box orbits in this
case is $b<c<a$.
Clearly, loop orbits obtain for $c<b$ and for $c>a$.

There is a simple geometric interpretation of the criterion
(\ref{eq:BLCriteron}) for box orbits and recession: it requires
that, if the two ellipses are drawn with a common centre, they
intersect each other.

\subsection*{Squaring the circle}

The domain of the the orbit
(\ref{eq:pertSHMx})--(\ref{eq:pertSHMy})
of the simple harmonic oscillator,
in the generic case of irrational $\delta^\prime$, 
is dense in a rectangular area.  If, for simplicity, we assume
$x_0=y_0$, the orbit covers a square.
It is not immediately obvious how this may be expressed in
terms of epi-elliptic motion. However, let us assume that
$\nu_1=\mu_2=0$, so that the elliptic components degenerate
into two orthogonal line segements and the solution
(\ref{eq:gensocmu})--(\ref{eq:gensocnu}) becomes
\begin{equation}
\mu = \mu_1\cos\beta_1(t-t_1) \,,
\qquad
\nu = \nu_2\sin\beta_2(t-t_2) \,.
\nonumber
\end{equation}
This is isomorphic to the solution 
(\ref{eq:pertSHMx})--(\ref{eq:pertSHMy}).
Thus, the circle is squared.







\end{document}